\documentclass[nofootinbib,12pt,reprint,longbibliography]{revtex4-2}
\usepackage[scr=boondoxo]{mathalpha}

\newcommand{\lie}[1]{\pounds_{#1}}
\newcommand{\odr}[2]{\frac{d #1}{d #2}}
\newcommand{\pdr}[2]{\frac{\partial #1}{\partial #2}}

\usepackage{graphicx}
\usepackage{epstopdf}
\usepackage{amsmath}
\usepackage{amssymb}
\usepackage{amsfonts}
\usepackage{amsthm}
\usepackage{mathtools}
\usepackage{color}
\usepackage{array}
\usepackage{verbatim}
\usepackage{url}
\usepackage{graphicx}
\usepackage{braket}
\usepackage{epsfig}

\usepackage[
colorlinks=true,
linkcolor=blue,
urlcolor=magenta,
filecolor=blue,
citecolor=red,
pdfstartview=FitV,
pdftitle={},
pdfauthor={},
pdfsubject={},
pdfkeywords={},
pdfpagemode={},
bookmarksopen=true
]{hyperref}
\usepackage{lipsum}
\usepackage{lipsum}
\usepackage{cleveref}
\crefname{figure}{FIG.}{figures}
\Crefname{figure}{Figure}{Figures}

\usepackage{caption, subcaption}

\def\frac#1#2{{#1\over #2}}

\def\be{\begin{equation}}
	\def\ee{\end{equation}}
\def\ba{\begin{eqnarray}}
	\def\ea{\end{eqnarray}}

\usepackage{titlesec}
\titleformat{\paragraph}[runin]
{\normalfont\scshape\bfseries}{\theparagraph}{1em}{}

\begin{document}
	
	\title{Slowly evolving horizons in Einstein gravity and beyond}

	\author{Ayon Tarafdar}\email{ayon.trf@gmail.com}
	\affiliation{Department of Physics, University of Calcutta,\\92, Acharya Prafulla Chandra Road, Kolkata - 700009, India }
	
	\author{Srijit Bhattacharjee}\email{srijuster@gmail.com}
	\affiliation{Indian Institute of Information Technology (IIIT), Allahabad, \\
		Deoghat, Jhalwa, Prayagraj- 211015, India}
	
	\date{\today}
	
	\begin{abstract}
	We study event horizon candidates for slowly evolving dynamical black holes in General Relativity and Einstein-Gauss-Bonnet (EGB) gravity. Such a type of horizon candidate has been termed as {\it slowly evolving null surface} (SENS). It signifies a near-equilibrium state of a dynamic black hole. We demonstrate the time evolution of such surfaces for three different metrics. First, we locate such a surface for a charged Vaidya metric and show that the parameter space of the black hole gets constrained to allow a physically admissible slowly evolving null surface. We then consider a supertranslated Vaidya solution that contains a non-spherical horizon and study the properties of the SENS. This spacetime generates a non-vanishing shear at the SENS due to the presence of the supertranslation field. The SENS for a spherically symmetric Vaidya-like solution in EGB gravity yields a bound on the accretion rate that depends on the size of the horizon.  We also show that the first and second laws of black hole mechanics can be established for these slowly evolving surfaces. 
	\end{abstract}

	\maketitle
	
\section{Introduction}\label{introduction}
Black Holes (BH) are arguably the most useful testing beds in the sky for uncovering innumerable facts regarding gravitational physics. The event horizon of a black hole is a null hypersurface that offers most of the interesting features of it. Further, the laws of black hole mechanics established on the event horizon intriguingly anticipate the laws of thermodynamics \cite{bardeen1973,hawking1975,wald2001}. This fascinating interconnection between quantum physics and the geometric laws of black hole mechanics is perhaps the strongest motivation behind the efforts of establishing a quantum theory of gravity. However, finding the location of the event horizon requires a complete knowledge of the future for a black hole spacetime. Moreover the laws of black hole mechanics, established on stationary event horizons,  are not always suited for practical situations where a black hole becomes dynamic due to its interaction with the surroundings. To understand the near-equilibrium situation or a scenario when a black hole geometry evolves, one needs an alternative set up where the response of a black hole's horizon can be studied without resorting to the global aspects. 

The non-equilibrium scenario of black holes (BH) has been addressed successfully in a quasi-local set-up where one focuses on \textit{trapped surfaces}. A trapped surface is a closed, spacelike codimension-2 surface with the property that all future-directed null normal congruences have decreasing cross-sectional area. The boundary of such a region serves as a useful way to characterize horizons from a geometric point of view rather than a causal one. Such a local definition of black hole horizons and corresponding laws of mechanics were first advocated by Hayward \cite{hayward1994} and extended by Ashtekar et al \cite{ashtekar2000, ashtekar2003, ashtekar2004}. The boundaries of such surfaces are usually regarded as {\it apparent horizons} or \textit{trapping horizons}. {\it Isolated horizons} and {\it dynamical horizons} were  proposed for a BH in equilibrium with its surroundings and one that is not, respectively \cite{ashtekar2004}. The laws of black hole mechanics have been successfully extended to these quasi-local horizons for general relativistic BH. These horizons are devoid of the teleological nature of the event horizon as they respond only when matter or radiation falls into them.  

The quasi-local trapping horizons are still not very useful in the case of asymptotically flat BH. \textit{Marginally trapped surfaces} for which the expansion of the outgoing null normal congruence vanishes - are situated within the event horizon and disconnected from the exterior. To avoid this difficulty one may rather choose a surface which is evolving and sufficiently close to the trapping horizon but not trapped. This set-up is somewhat related to the membrane paradigm that was developed for the sake of probing the near horizon physics of BH \cite{thorne1986}. 

 Among the different types of trapping horizons the {\it future outer trapping horizon} (FOTH) turns out to be important as it provides an equivalent local description of an event horizon \cite{hayward1994}. A slowly evolving FOTH was introduced by Booth and Fairhurst in \cite{booth2004,booth2007} that can emulate near-equilibrium behaviour. In this article we study the trapping horizons of BH in Einstein gravity and beyond. Particularly, we focus on locating a \textit{slowly evolving null surface} (SENS), introduced in \cite{booth2010, booth2013}, that behaves similar to a weakly perturbed event horizon. The slowly evolving horizons are not only useful to accommodate initial and final states that are not infinitesimally separated, but they also provide a simple and effective way to characterize near-equilibrium situations. Studying these horizons may shed useful insight on the near-equilibrium phases of colliding BH in the early or very late stage of merger \cite{dreyer2003a}.



The paper is organised as follows: in the next section, we describe the notation and the set-up to be used throughout the paper. In section III, we introduce the FOTH of a charged Vaidya metric and discuss the properties of it. We also introduce the SENS to the reader. Location and properties of the SENS are detailed thereafter. In the next section we study the SENS of a supertranslated Vaidya black hole having a non-spherical horizon. However, the spacetime still has a spherical topology. The location of slowly evolving horizons in this case are dependent on the angular variable $\theta$. The consequences are discussed with plots. The next section contains the study of SENS for a Vaidya black hole in EGB gravity in 5 dimensions. Next, the first and second law of black hole mechanics for the SENS are introduced. The validity of these laws for the EGB BH is discussed. Finally, we conclude by highlighting notable outcomes of our study and indicate some future prospects.

\section{Notation and set-up}
\label{sec:nsetup}
The mathematics of perturbatively constructing the spacetime around a foliated, codimension-1 hypersurface (in this case the FOTH) is well-known and our immediate references are \cite{booth2013,booth2007}. In this section, we just highlight the key geometric structures and results we need.

A manifold $\mathcal{M}$ of dimension $d+1$ is endowed with a metric $g_{ab}$ and its corresponding connection $\nabla_a$. We denote the spacetime indices by lower case initial Latin letters $(a,b,\cdots)$.\ The quantities on $d$-dimensional hypersurfaces are denoted by $(i,j,k,\cdots)$ and capital letters $(A,B,\cdots)$ are used to denote quantities on $d-1$ dimensional surfaces. We use $(-,+,+,\cdots, +)$ signature for the metric. We also set $G = c = 1$.

A \textit{future outer trapping horizon} (FOTH) is defined as a hypersurface $\Sigma$ of dimension $d$ with the following properties: i) They are foliated by a family of $d-1$ dimensional spacelike surfaces $S_v$ with future-directed null normals $l$ (outgoing) and $n$ (ingoing) on it. Here $v$ is just a coordinate that labels the spatial surfaces. The expansion w.r.t the null vector $l$ vanishes i.e.\ $\Theta_{(l)}=0$ on FOTH. ii) The expansion w.r.t the other null normal $n$ is negative, $\Theta_{(n)} <0 $. iii)  $\mathcal{\delta}_{n}\Theta_{(l)}<0$, where $\delta_{n}$ is the Lie derivative $\lie{n}$ pulled-back on the $d-1$ surface. 

The null normals are cross-normalized as $l\cdot n=-1$. Now we define a vector field along which the leaves of the foliation ($S_v$) evolve. It is tangent to the horizon but normal to the $d-1$ surface as:
\begin{equation}\label{eq:tangent-vec}
    \mathcal{V}^a=l^a-Cn^a,
\end{equation}

where $C$ is a scalar field. This choice allows us to fix the scaling freedom of the null vectors. Under the rescalings:
\[
    l^a\to h(v)l^a,\quad
    n^a\to n^a/h(v),\quad
    C\to h(v)^2 C,
\]
we have $\mathcal{V}^a\to h(v)\mathcal{V}^a$, which amounts to a relabelling of the foliation leaves. The norm of this vector is $\mathcal{V}^a\mathcal{V}_a=2C$. When the dominant energy condition holds, it can be shown using $\mathcal{\delta}_{n}\Theta_{(l)}<0$ that $C\geq 0$.
A FOTH is spacelike for $C>0$ and null for $C=0$. We will not consider the timelike  ($C<0$) case here. The $d-1$ dimensional metric is given by $q_{ab}=g_{ab}+l_an_b+l_bn_a$. The pulled-back metric on this surface is given by $q_{AB}=e^a_Ae^b_B\,g_{ab}$, where we denote $e$ as the pull-back maps. The extrinsic curvatures of this codimension-2 surface are defined as 
\begin{equation}
    \mathcal{K}^{(l)}_{AB}=e^a_Ae^b_B\nabla_al_b, \quad \mathcal{K}^{(n)}_{AB}=e^a_Ae^b_B\nabla_an_b.
\end{equation}
The expansion $\Theta_{(l)/(n)}$ and shear $\sigma^{(l)/(n)}_{AB}$ of the null congruences are obtained as the trace and trace-free parts of these extrinsic curvatures. The evolution of the expansion along the null congruences is given by the Raychaudhuri equation and they indicate how the geometry of the $d-1$ surface evolves along the null direction. 
\begin{equation}
    \lie{l}\Theta_{(l)}=\kappa_{(l)}\Theta_{(l)}-\frac{\Theta_{(l)}^2}{d-1}-\sigma_{AB}^{(l)}\,\sigma^{AB}_{(l)}-\mathcal{R}_{ab}l^a l^b,
\label{RE}\end{equation}
where $\kappa_{(l)}$ is the dynamical surface gravity defined as 
\[\kappa_{(l)}=-l^an_b\nabla_a l^b.\]

A similar expression for $\Theta_{(n)}$ can also be obtained. Now, on an FOTH $\delta_{\mathcal{V}}\Theta_{(l)}=0$. Using this, we can obtain a second order differential equation for $C$ \cite{booth2007}:
\begin{equation}
    d^2 C - 2\omega^A d_A C - \delta_l\Theta_{(l)} + C\delta_n\Theta_{(l)} = 0.
\end{equation}
where $d_A$ is the covariant derivative of the $(d-1)$-surface. $\omega_A = -e^a_A n_b\nabla_a l^b$ is the connection on the normal bundle.  
For spherical symmetry, the ($d-1$)-surface derivatives drop out and we're left with an explicit expression for $C$.\\

{\bf Slowly evolving horizon:} When one refers to a slowly evolving horizon, one commonly refers to the FOTH, satisfying the slowly evolving conditions. These conditions are determined from the intrinsic and extrinsic geometry of the FOTH, discussed at length in \cite{booth2013}. 

A slowly evolving FOTH can be characterized as a hypersurface that is perturbatively non-isolated with the surroundings. One of the key slowness conditions is that the evolution parameter $C$ has to be small. To obtain the SENS, we perturbatively expand around the FOTH along radial null geodesics normal to the spatial codimension-2 surface. In this regime, the SENS coincides with the event horizon of an asymptotically stationary spacetime \cite{booth2013}. Hence it can also be termed as a `slowly evolving horizon'. For a BH in an equilibrium state, the SENS naturally coincides with the FOTH, which then becomes null.\\

{\bf Slowly Evolving Null Surface:}  We now formally introduce the conditions an SENS must satisfy. A section of a $d$-dimensional null surface with tangent vector $l^a$ and a characteristic scale $R_{\Sigma}$ (typically given by the areal radius of the horizon) is a slowly evolving null surface (SENS) if \cite{booth2013}:
\begin{enumerate}
    \item[1.] \hfill$\begin{aligned}[t]
 \frac{1}{d-1}\Theta_{(l)}^2\ll\big(\sigma_{AB}^{(l)}\,\sigma^{AB}_{(l)}+\mathcal{R}_{ab}l^a l^b\big),
\end{aligned}$\hfill\null
\end{enumerate}
and $l^a$ can be scaled so that $\kappa_{(l)}$ is of order $1/R_{\Sigma}$ so that:
\begin{enumerate}
\item[2.] \hfill$\begin{aligned}[t]
\lie{l}\Theta_{(l)}\ll \kappa_{(l)}\Theta_{(l)}.
\label{sensc-2}\end{aligned}$\hfill\null
\end{enumerate}

These conditions allow us to treat the SENS as a surface that mimics a near-equilibrium situation of a weakly perturbed event horizon or isolated horizon. 

We start with a discussion of the charged Vaidya metric and use it to introduce our procedure and additional useful notation.

\section{Charged Vaidya metric}\label{sec:charged-vaidya-metric}
The charged Vaidya metric or the Vaidya-Reissner-Nordström (VRN) metric in 4-d is a spherically symmetric metric which in ingoing Eddington-Finkelstein coordinates is \cite{bonnor1970}:
\begin{align}
ds^2 &= -\triangle(v,r)dv^2 + 2dvdr + q_{AB}dx^Adx^B,\\ 
\triangle(v,r) &= 1-\frac{2m(v)}{r}+\frac{q(v)^2}{r^2}\cdot
\end{align}

The spatial 2-surface is described by the metric of a sphere:
\[
q_{AB}dx^Adx^B=r^2(d\theta^2+\sin^2 \theta\;d \phi^2).
\]
To determine the trapping horizon, we focus on the outgoing and ingoing radial null geodesics, the tangent vectors to which are $l^a$ and $n^a$ respectively. These are
\begin{equation}
    l^a\coloneqq\left(1,\frac{\triangle}{2},0,0\right),\quad n^a\coloneqq(0,-1,0,0).
\end{equation}
The expansion of these null congruences are:
\[
\Theta_{(l)}=\frac{\triangle}{r},\quad \Theta_{(n)}=- \frac{2}{r}.
\]
The expansion of the ingoing null normal, as expected, is everywhere negative.
\subsubsection{The energy condition}\label{sec:the-energy-condition-vrn}
The stress-energy tensor for this metric looks like 
\[
T_{ab} = T^{(em)}_{ab} + T^{(ex)}_{ab}
\]
where the $T^{(em)}_{ab}$ is the electromagnetic part and the extra term $T^{(ex)}_{ab}$ is
\begin{equation}
    8\pi T^{(ex)}_{ab} = \frac{2}{r^3}(\dot m r - q\dot q)\,\delta_a^v \delta_b^v
\end{equation}
(we've dropped the function arguments for brevity).

This satisfies the null energy condition (NEC) $T_{ab}l^a l^b\geq0$ for the outgoing null vector $l^a$ if
\[
r\,\dot m(v) - q(v)\dot q(v) \geq 0\quad\text{or}\quad r \geq \frac{q(v)\dot q(v)}{\dot m(v)},
\]
where we can term $r_{cs} \coloneqq q\dot q/\dot m$ as a ``critical surface". If this critical surface exists $(r_{cs} > 0)$, within this surface, the NEC will be violated. In our case, the critical surface needs to be within the FOTH. Otherwise, even for accretion of charged null dust, the FOTH radius may decrease  \cite{mishra2019}. Denoting it by $r_+$, this means $r_+\geq r_{cs}$.
\subsection{Locating the FOTH}\label{sec:locating-the-foth-vrn}
The trapping horizon is located where $\Theta_{(l)}=0=\triangle$, which gives two roots:
\begin{equation}
    r_{\pm}(v)= m(v)\,\pm\sqrt{m(v)^2-q(v)^2}.
\end{equation}
We're interested in the larger root $r_+(v)$, which is our ``outer" horizon, i.e. on this hypersurface, $\lie{n}\Theta_{(l)}<0$. Computing this quantity for this metric we get
\[
\lie{n}\Theta_{(l)}=-\frac{2}{r_+^3}\sqrt{m^2-q^2}.
\]
Since $m>q$ and $r_+ >0$, the RHS is always negative. Moreover, since $\Theta_{(n)}<0$, it is of future type.
Thus $r_+(v)$ locates our future outer trapping horizon (FOTH). We'll use this notation for the FOTH throughout our paper. Considering $\Phi\coloneqq r-r_+(v)$ as the equation of this hypersurface, the norm of its normal is
\[
g^{ab}\Phi_{,a}\Phi_{,b}=\frac{2}{\sqrt{m^2-q^2}}(q\dot q-\dot m r_+).
\]
When the null energy condition is strictly satisfied, we have $\dot{m}r_+ > q\dot{q}$. Thus the RHS is always negative and the FOTH is spacelike.

\begin{figure*}
\begin{centering}
\includegraphics[width=\linewidth]{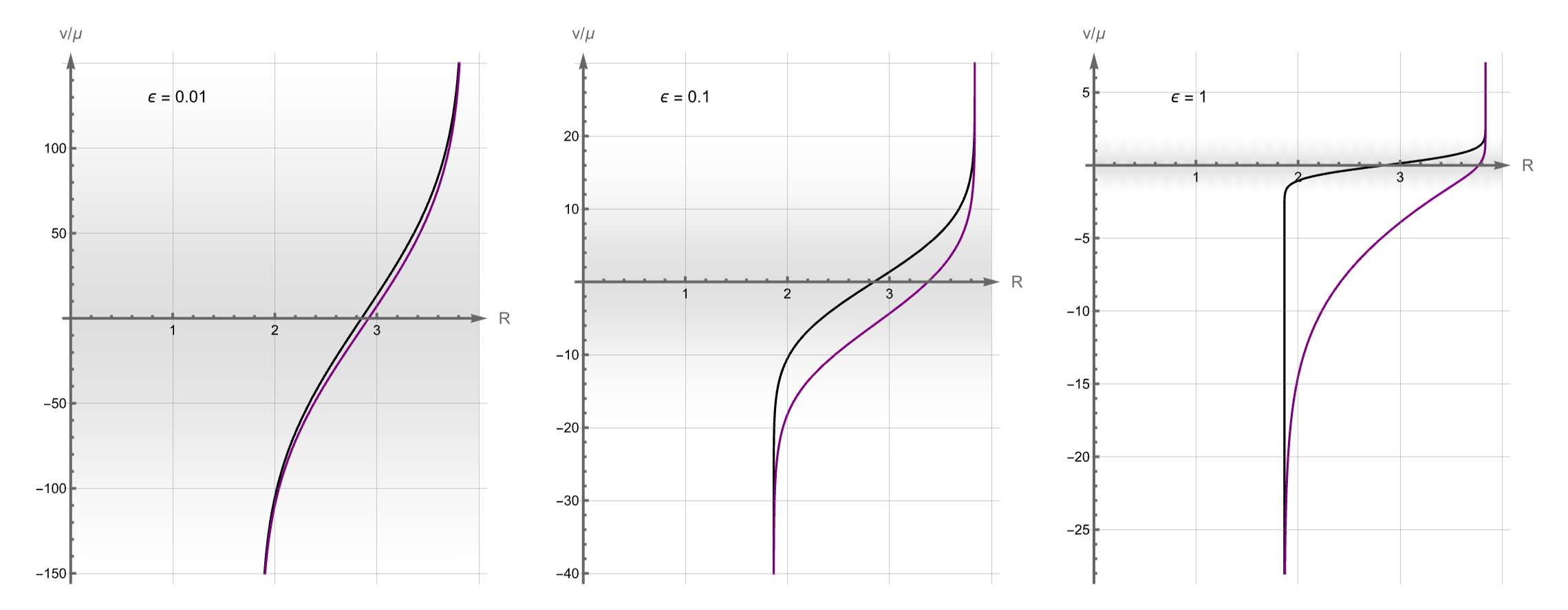}
\caption{Plotting the FOTH (black) and numerical solution of the ``actual" event horizon (purple) for different values of $\epsilon$. The gray shading indicates the mass flux. The y-axis has $v/\mu$ instead of $V$ to demonstrate the effect of the length scale.}
\label{fig:vrn-foth-eh}
\end{centering}
\end{figure*}

\subsection{Locating the SENS}\label{sec:locating-the-sens-vrn}
To locate the SENS we wish to find a null surface that is obtained as a perturbative expansion around the FOTH. For this we introduce a mass scale $\mu$ and a length scale $L$ and switch to dimensionless quantities. This allows us to study a range of spacetimes in a similar fashion \cite{booth2010}. We do the following transformations:
\[
r\to \mu R,\;\;
v\to L \mu V,\;\;
m(v)\to \mu M(V),\;\;
q(v)\to \mu Q(V).
\]
Since $v$ is our time coordinate, $L$ acts more like a time scale in our study. We will particularly be interested in $L\gg 1$ regimes, where our spacetime evolves ``slowly". We will also often use $\epsilon\equiv 1/L$. It is convenient to also transform the line element:
\begin{equation}
    \mu^{-2}ds^2 = -\triangle(V,R)L^2dV^2 + 2L\,dVdR + R^2 d\Omega^2
.\end{equation}
Notice that $\mu^2$ comes out as an overall factor and hence is not that important. No factors of $L$ appear within $\triangle(V,R)$. Extra factors of $L$ will appear whenever there is a derivative w.r.t $v$. For instance:
\[
\dot m(v)=\odr{m(v)}{v}\to \frac{\mu}{\mu L}\odr{M(V)}{V}= \frac{1}{L}\dot M(V).
\]
The crucial difference in our analysis with \cite{booth2013} is we have only extracted the time scale out of the problem (in $L$ or equivalently, $\epsilon$) - which naturally appears with time rate-of-change quantities and decides whether the spacetime is dynamical or not in the first place, whereas they expand in terms of the parameter $C$, which perhaps incorporates the intrinsic scale of the spacetime more naturally. In the Vaidya case, extracting the time scale is enough, since that is the only relevant scale. 

For numerical work, we choose the following functions (automatically satisfying the NEC):
\begin{equation}
\begin{split}
M(V)&= \frac{3}{2}+ \frac{1}{2}\text{erf}{(V)},\\
Q(V)&= \frac{1}{2}M(V)^{2/3}.
\end{split}
\end{equation}
Here $'\text{erf}'$ denotes the error function.  We now posit the location of the null surface as an expansion in $\epsilon$ about $R_+(V)$:
\begin{equation}
    R_n(V)\coloneqq R_+(V)\left[1+\sum_{i=1}^{N}\epsilon^i A_i(V)\right],
\end{equation}

where $N$ is the order at which we truncate the series. $A_i$ is the perturbation coefficient at the $i^{th}$ order. The radial null geodesics satisfy the nullity condition
\begin{equation}
    \odr{r}{v}= \frac{1}{2}\triangle(v,r),
\end{equation}
which in dimensionless quantities becomes
\begin{equation}\label{eq:dimless-nullity-vrn}
    \epsilon\odr{R}{V}= \frac{1}{2}\triangle(V,R).
\end{equation}

We now solve this equation order-by-order in $\epsilon$. To proceed, we must make some assumptions:
\begin{enumerate}
\item 
$\epsilon$ is small, likely $\epsilon\ll 1$.
\item 
The following hierarchy of derivatives holds: $R_+ \gtrsim \dot R_+ \gtrsim |\ddot R_+|\,\dots$
\end{enumerate}
The origin of the second assumption can be tied to the area of the horizon. Spacetimes for which we can write the horizon area $A\propto R_+^2$, the derivatives go as $\dot A\propto R_+ \dot R_+, \ddot A \propto (\dot{R}_+^2+\ddot R_+)\,\dots$ Thus the second condition requires us to make an assumption that the successive derivatives of the area of the horizon are of the order of the area itself. In the Vaidya case, this simply reduces to $M \geq \dot M \geq \ddot M\dots\,$, a condition on the derivatives of the mass function itself. In case of VRN or more complicated spacetimes, this might not be the case.

We solved \cref{eq:dimless-nullity-vrn} upto $N=2$, but extending it to higher orders is straightforward. The first of these coefficients is
\begin{equation}
    A_1\equiv R_+\frac{\dot M R_+ - Q\dot Q}{M^2-Q^2}.
\end{equation}
Taking a close look at it, we again see that this coefficient is always positive for the case in hand: no violation of NEC and for a non-extremal VRN BH.  Since this is the leading order correction, this ensures that the perturbed surface ``encloses" the FOTH: i.e.\ on every constant-$V$ slice, its radial coordinate location is everywhere greater than that of the FOTH.

If we wish to write the perturbed expression in terms of the original mass and charge functions, we can absorb $\epsilon$ and $\mu$ back wherever appropriate to yield, straightforwardly
\begin{align}
r_n(v)&\coloneqq r_+(v)\left[1+\sum_{i=1}^{N}a_i(v)\right]\\
a_1 &= r_+\frac{\dot m r_+ - q\dot q}{m^2-q^2}
\end{align}
However, to enforce the original restriction of $\epsilon\ll 1$, the derivatives of the radial location now follow the hierarchy:
\[
r_+\gg \dot r_+\gg |\ddot r_+|\dots
\]
\begin{figure}
\begin{centering}
\includegraphics[width=\linewidth]{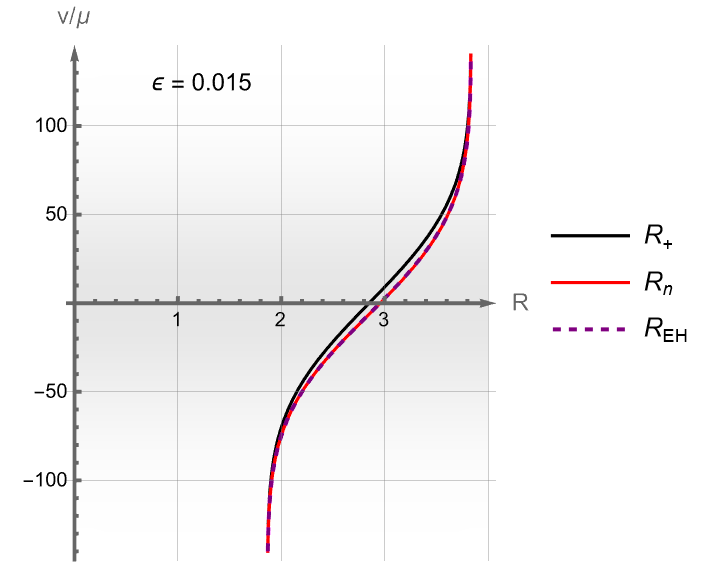}
\caption{With the null surface (red) in the slowly evolving regime, compared with the actual event horizon (purple, dashed). At least visually, they seem to coincide well.}
\label{fig:vrn-seh}
\end{centering}
\end{figure}
\subsubsection{Verifying the slowly evolving conditions}\label{sec:verifying-slowly-evolving-conditions-vrn}

Since the SENS has been generated by outgoing radial null geodesics, we can find the expansion of this congruence at $R_n$, to leading order in $\epsilon$:
\[
\Theta_{(l)}=2\epsilon\frac{R_+ \dot M - Q\dot Q}{\mu R_+\sqrt{M^2-Q^2}}=\frac{2\epsilon \dot R_+ }{\mu R_+}
\]
which is positive, given our conditions. Thus this surface is untrapped.

The slowly evolving conditions for a spherically symmetric spacetime (and slicing) do not involve shear. The remaining quantities that appear in the condition, all at leading order are:
\begin{equation}
\begin{gathered}
\kappa_{(l)} = \frac{\sqrt{M^2-Q^2}}{\mu R_+^2},\quad
G_{ab}l^a l^b = 2\epsilon\frac{R_+ \dot M - Q\dot Q}{\mu^2 R_+^3},\\ \lie{l}\Theta_{(l)}=\frac{2\epsilon^2}{\mu^2 R_+^2}\Big(R_+ \ddot R_+ -\dot R_+^2\Big),
\end{gathered}
\end{equation}
where the expression for $\lie{l}\Theta_{(l)}$ has been written in terms of the FOTH radius and its derivatives. Here we are just interested in the leading order of $\epsilon$ at which these quantities appear, and possibly their signature. Then for $\epsilon\ll 1$ and the assumptions previously made, the slowly evolving conditions are achieved. This null surface obtained, in the slowly evolving regime, coincides with the event horizon of this spacetime \cite{booth2013}.

However, as $q/m\to 1$, the scenario is a little involved. Then $\kappa_{(l)},G_{ab}l^a l^b\to 0$ but $\Theta_{(l)}$ and $\lie{l}\Theta_{(l)}$ remain finite. Clearly, we need to find a bound on $q/m$. To do so, we look at the very origin of the definition of a slowly evolving horizon.
Using the evolution vector in \cref{eq:tangent-vec}, one can define a slowness parameter\footnote{In\cite{booth2007,booth2013}, $\beta$ is $\varepsilon$. Our use of $\epsilon$ is different, as noted earlier.}
\[
\beta\coloneqq \sqrt{\left|\frac{1}{2}C \Theta_{(n)}^2 R_{\Sigma}^2\right|}\ll 1
\]
evaluated at the FOTH. This definition does not work for the SENS since $C = 0$ on it. In all of our examples, $\Theta_{(n)}=-2/r$. Evaluated at the FOTH, the parameter becomes
\[
\beta=\sqrt{2C}\ll 1\implies C\ll \frac{1}{2}
\]
\begin{figure*}
\begin{centering}
\includegraphics[width=\linewidth]{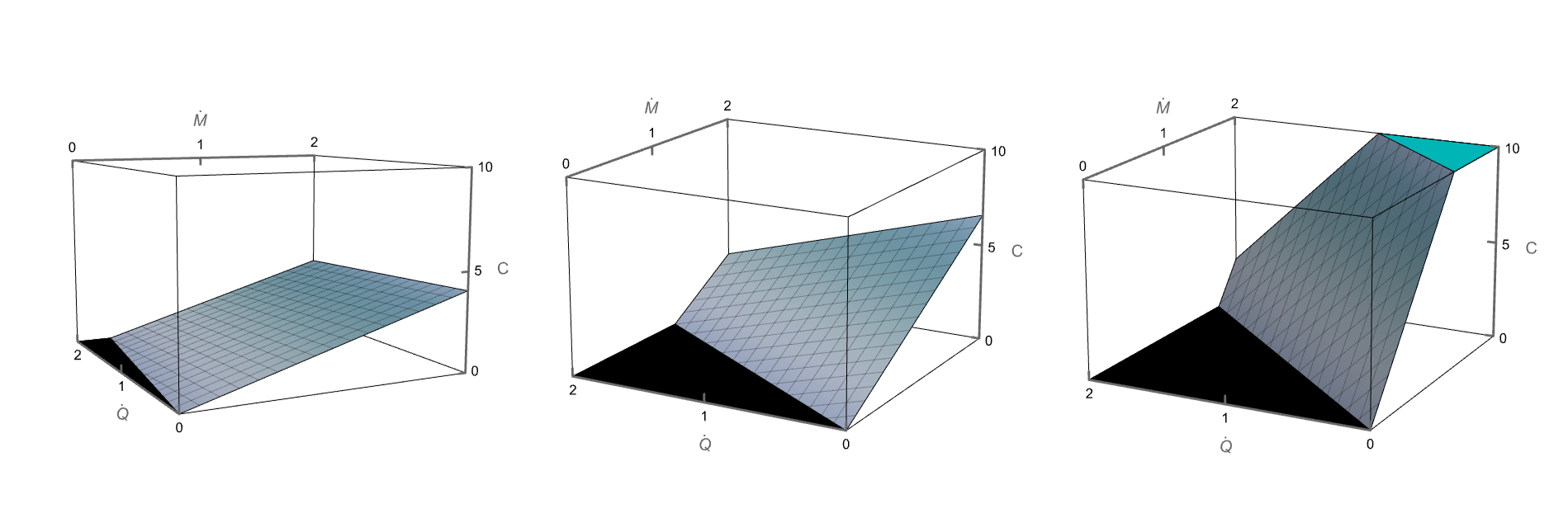}
\caption{The variation of the (scaled) evolution parameter on a constant-V slice with $\dot M$ and $\dot Q$ for different values of $Q/M$. The clipped black regions are where NEC is violated. The value of $C$ should stay at the order of unity.}
\label{fig:vrn-param-extremal}
\end{centering}
\end{figure*}
If this condition is imposed on the various expressions obtained, it ensures that we stick to the slowly evolving regime.
We can use this to obtain the required bound. In the Vaidya case, $C = 2\dot m$ and thus $\dot m$ being small ensures that the horizon is slowly evolving. In our case, however,
\begin{equation}
    C = \frac{r_+\dot m - q\dot q}{\sqrt{m^2-q^2}}\cdot
\end{equation}
In terms of dimensionless quantities the slowly evolving condition reads
\[
2\epsilon\dot M\frac{R_+ - R_{cs}}{\sqrt{M^2-Q^2}}\ll 1
\]
Now we note that $\epsilon\ll 1$. Thus the terms that appear as a product with $\epsilon$ must be at most as large as unity. This gives us an inequality:
\[
\sqrt{1-\frac{Q^2}{M^2}}\geq 2\left(\frac{\dot M}{M}\right)(R_+ - R_{cs}),
\]
which seemingly gives us a lower bound on how close can $Q$ be to $M$. But this is not quite true, since both sides contain $Q$. For the analysis, we look at a particular value of $M$ and change the value of $Q$.
A special case would be when $Q=const.$, implying $R_{cs}=0$.\footnote{This would also work if $R_{cs}$ is negligible compared to $R_+$.} Then the condition simplifies further:
\[
\frac{\sqrt{1-Q^2/M^2}}{1+\sqrt{1-Q^2/M^2}}\geq 2\dot M
\]
which can be thought of as a bound on $\dot M$ that depends on the proximity to extremality. As $Q$ approaches $M$, $\dot M$ is required to be increasingly smaller for the horizon to be slowly evolving.

Alternatively, $R_{cs}>0$ approaches $R_+$. This is possible to achieve just by an appropriate choice of $\dot Q$ and $\dot M$. In this regime, we see that it becomes easier to satisfy the inequality. In fact, in the limit $R_{cs}\to R$, $Q$ can be arbitrarily close to $M$. These cases are demonstrated in \cref{fig:vrn-param-extremal}.

Further, we know that the surface gravity $\kappa_{(l)}$ for the extremal BH is indeed zero. Therefore, for the expansion about the slowly evolving FOTH to work, $\kappa_{(l)}$ must be restricted to a particular length scale ($\sim 1/R_{\Sigma}$). As $\kappa_{(l)}$ approaches zero, the coefficients of expansion also become arbitrarily large. However, with a bound put on how close can $Q/M$ be to unity and/or ensuring that the hierarchy of derivatives hold, the expected results for slowly evolving horizons follow.

\subsubsection{Conformal Killing horizons}\label{sec:conformal-killing-horizon}
Interestingly Vaidya BH admit a conformal Killing horizon on the SENS for linear mass functions \cite{nielsen2014,nielsen2018}. The existence of such conformal mappings allows us to relate properties of dynamical spacetimes to static ones. We explore conformal Killing horizons for the charged Vaidya metric in this section. Considering the vector
\begin{equation}
k^a \coloneqq (v,r,0,0)
\end{equation}
which satisfies the conformal Killing equation
\[
\nabla_{(a}k_{b)}= 2g_{ab}
\]
if the following relation holds:
\[
\frac{1}{r^2}[q(q-v\dot q)-r(m-v\dot m)]=0.
\]
Clearly this is satisfied if $q=\dot q v$ and $m=\dot m v$, which yield linear solutions for both the mass and the charge functions $m(v)=c_m v$ and $q(v)=c_q v$. Finally, the norm of this vector is
\[
k^a k_a = 2rv - v^2\triangle.
\]
If we evaluate this norm at our slowly evolving horizon, putting $r=r_n$, we find that it becomes zero (upto the order which it was evaluated at). Thus for linear mass functions, the SENS is also a conformal Killing horizon. We have chosen one of the simplest possible ansatz for the conformal Killing vector $k^a$. For other choices which satisfy the general conformal Killing equation
\[
\nabla_{(a}k_{b)}= \Psi^2g_{ab},
\]
one might be able to choose more involved mass functions.

\section{Supertranslated Vaidya metric}\label{sec:supertranslated-vaidya-metric}
The supertranslated Vaidya metric (STV) is a departure from the comfort of spherical symmetry. Slowly evolving horizons in spherically asymmetric spacetimes have been investigated in \cite{kavanagh2006}, where a black hole of constant mass moving in a slowly changing tidal field was treated as a perturbation. This was largely concerned with the analysis of the FOTH, which turned out to be spherical upto the octupole order. In the STV case, the metric is obtained after imparting a linear supertranslation hair to a Vaidya black hole with an appropriate stress-energy tensor \cite{strominger2018, chu2018}. The FOTH in this case is not spherically symmetric, but it has spherical topology. This metric is described in terms of the following components \cite{sarkar2022}:
\begin{equation}
\begin{split}
ds^2 &= -g_{vv}dv^2 + 2dv dr + g_{v\theta}dv d\theta\\
&+ r^2 (\tilde{g}_{\theta \theta}d\theta^2 + \tilde{g}_{\phi\phi}\sin^2{\theta}\,d\phi^2)\\
g_{vv}&=V-\left(\frac{m}{r^2}f''+\frac{m}{r^2}f'\cot{\theta}+\frac{2\dot m}{r}f\right),\\
g_{v\theta}&=f'\csc^2{\theta}-2Vf'-f''\cot{\theta}-f''',\\
\tilde{g}_{\theta\theta}&=1+\frac{1}{r}(f''-f'\cot{\theta}),\\
\tilde{g}_{\phi\phi}&=1-\frac{1}{r}(f''-f'\cot{\theta}),\\
\end{split}
\end{equation}
where
$V \coloneqq (1-2m(v)/r)$ is the Vaidya mass function and
$f(\theta)$ is the supertranslation field, an arbitrary function of the angular polar coordinate.
We restrict to only linear orders in $f$ and its derivatives.\footnote{Thus, when we say we discard $O(f^2)$ terms, we're discarding $O(f'^2),O(f''^2)\dots$ as well.} Thus this field only contributes as a correction to the spherically symmetric Vaidya metric, especially when $f\ll m$, as considered in \cite{chu2018}. 


The tangent vectors to the ingoing and outgoing radial null geodesics are:
\begin{equation}
l^a=\left(1,\frac{1}{2}g_{vv},0,0\right),\quad
n^a=(0,-1,0,0).
\end{equation}
Now we can find the expansion of the corresponding null congruences:
\[
\Theta_{(l)}=\frac{g_{vv}}{r}+O(f^2),\quad
\Theta_{(n)}=-\frac{2}{r}+O(f^2).
\]
\subsubsection{The energy condition}\label{sec:the-energy-condition-stv}
We can compute $G_{ab}l^a l^b$ to obtain the constraint that would be imposed in order to satisfy the null energy condition for this metric. Then $G_{ab}l^a l^b\geq 0$ implies
\[
(2\dot m +\ddot m f)r+2\dot m(f'\cot{\theta}+f'')\geq 0
\]
which would imply the existence of a critical surface
\[
r_{cs}\coloneqq \left(1+\frac{\ddot m f}{2\dot m}\right)^{-1}\bigg(f'\cot{\theta}+f''\bigg)
\]
which to linear order in $f$ is just
\begin{equation}
r_{cs}=f'\cot{\theta}+f''.
\end{equation}
Thus given a supertranslation field, the critical surface exists when $r_{cs}>0$. If this critical surface remains within the FOTH, i.e. $r_{cs}\leq r_+$, no issues should manifest. When we will be investigating slowly evolving horizons, $\dot m$ is considered to be small and this condition reduces to
\[
f' \cot{\theta} + f''\leq 4m.
\]
\begin{figure*}
\begin{centering}
\includegraphics[width=\linewidth]{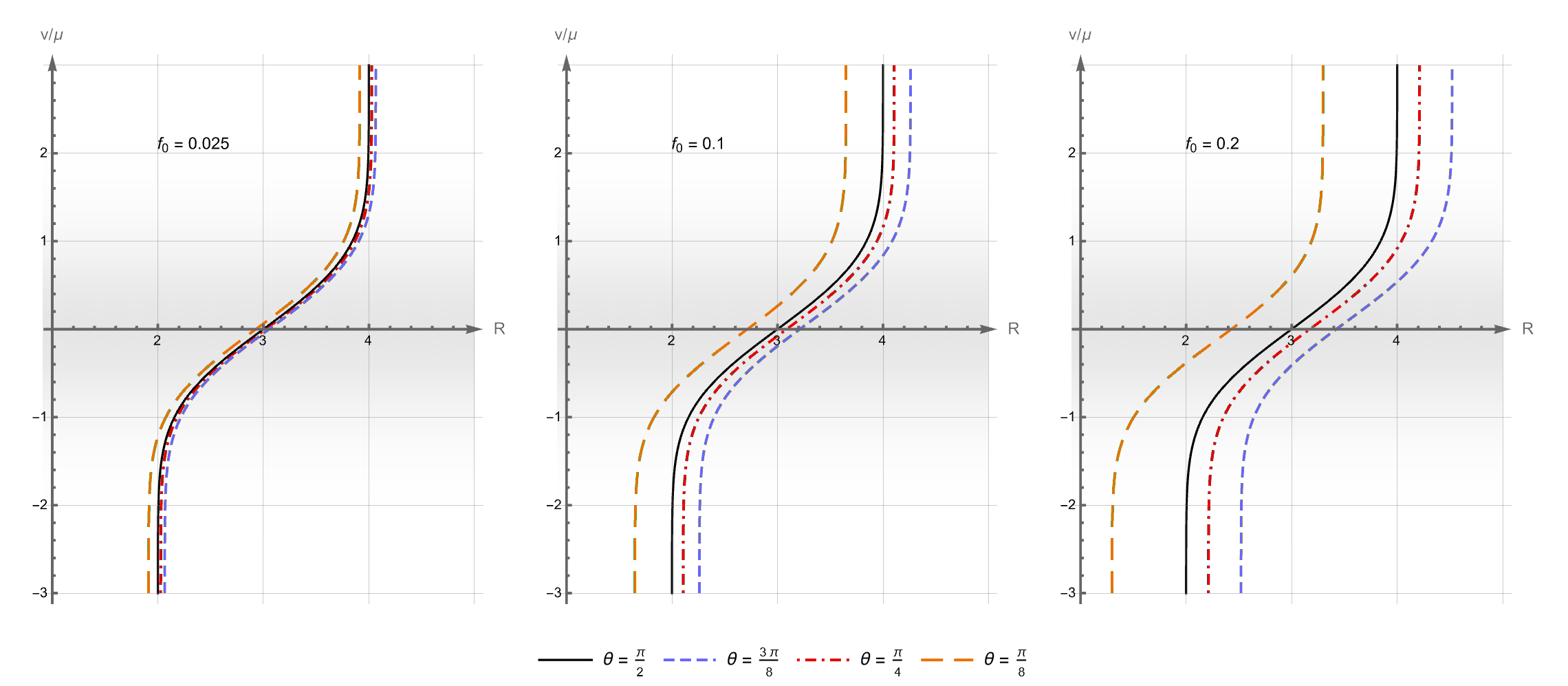}
\caption{Effect of changing values of $\theta$ and $f_0$ on the FOTH. As we stray beyond $\theta=\pi/4$, the FOTH deviates significantly compared to other values of $\theta$. As we move to $f_0=0.2$, corrections due to $f(\theta)$ are no more small compared to $m(v)$ (note, we aren't in the slowly evolving regime yet).}
\label{fig:stv-foth}
\end{centering}
\end{figure*}
\subsection{Locating the FOTH}\label{sec:locating-the-foth-stv}
Discarding $O(f^2)$ terms and setting $\Theta_{(l)}=0$, we see that the FOTH is located at
\begin{equation}
    r_+ = 2m + 2\dot m f + \frac{1}{2}(f' \cot{\theta} + f'')
\end{equation}
which, to be a physically reasonable solution, $2m$ has to be larger than the remaining terms. In the slowly evolving regime, this condition becomes $f' \cot{\theta} + f''> -4m$. That this is indeed the outer solution can be verified by checking $\lie{n}\Theta_{(l)}<0$:
\[
\lie{n}\Theta_{(l)} = \frac{1}{4m^{3}}\left[-m+2\dot mf+ \frac{1}{4}(f' \cot{\theta} + f'')\right]
\]
Under the assumption that the derivatives of $m(v)$ are comparable in magnitude to itself, this expression is always negative, since $f$ is assumed to be small anyway. It is worth noticing that in the slowly evolving regime, this is equivalent to satisfying the NEC.

However, we notice that the coordinate divergence due to the presence of the $\cot{\theta}$ factor is potentially problematic, since the term involving $f'$ does not remain small as $\theta\to 0$ or $\theta\to \pi$. Thus we should restrict ourselves to values of $\theta$ such that $\cot{\theta}\sim 1$. A reasonable way to obtain a bound on $\theta$ in the slowly evolving regime is to consider the two conditions just stated together:
\begin{equation}\label{eq:stv-theta-bound}
-1 < \frac{1}{4m}(f' \cot{\theta} + f'')\leq 1.
\end{equation}

\subsection{Locating the SENS}\label{sec:locating-the-sens-stv}
To find the null horizon candidate, we introduce dimensionless quantities again:
\[
r\to \mu R,\quad
v\to L \mu V,\quad
m(v)\to \mu M(V),\quad
f(\theta)\to \mu \mathscr{f}(\theta)
\]
The metric function $g_{vv}$ now picks up a factor of $\epsilon$ due to the presence of $\dot m(v)$:
\[
g_{vv}=V-\left(\frac{M}{R^2}\mathscr{f}''+\frac{M}{R^2}\mathscr{f}'\cot{\theta}+\epsilon\frac{2\dot M}{R}\mathscr{f}\right)
\]
Previously made comments regarding this approach on the hierarchy of derivatives remain unchanged. Apart from the previously chosen mass function, the supertranslation field chosen for numerical work is:
\begin{equation}
    \mathscr{f}=\mathscr{f}_0 P_3(\cos{\theta})
\end{equation}
where $\mathscr{f}_0$ controls the magnitude of the field and $P_3$ is the 3rd order Legendre function.

Now the perturbative expansion around the FOTH:
\begin{equation}
    R_n(V,\theta)\coloneqq R_+(V,\theta)\left[1+\sum_{i=1}^{N}\epsilon^i A_i(V,\theta)\right]
\end{equation}
with coefficients that now depend on $\theta$ as well. Again, we will solve upto $N=2$.

Now we consider radial null geodesics again, which satisfy this nullity condition
\[
\epsilon\pdr{R}{V} = \frac{1}{2}g_{vv}(V,R,\theta)
\]
and solve order-by-order using our ansatz. Then the coefficients obtained are
\begin{equation}
\begin{split}
A_1 &= 4\dot M-\frac{4\dot M}{M}(\mathscr{f}'\cot{\theta}+\mathscr{f''})\\
A_2 &= 16(2\dot M^2+M\ddot M)+4\mathscr{f}\ddot M\\
&-4\bigg(\mathscr{f}'\cot{\theta}+\mathscr{f''}\bigg)\left(\frac{2\dot M^2}{M}+\ddot M\right)
\end{split}
\end{equation}
where we've written the terms in a suggestive way by keeping the supertranslation-induced terms separate. Setting $\mathscr{f}=0$ reduces these coefficients to the ones obtained in the Vaidya case.

On the question of these corrections being positive, for the SENS to ``encapsulate" the FOTH, we must keep in mind the restrictions imposed on the possible values of $\theta$. It is possible to find values of $\theta$ for which the SENS ``dips within" the FOTH. The reasonable and safe resolution to this is choosing $f\ll m$ which automatically respects (\ref{eq:stv-theta-bound}).

A few comments about our approach is in order. We have chosen to evolve the FOTH, a hypersurface which is not spherically symmetric along the radial null vector, $l^a$. In the previous case, this vector was normal to both the FOTH and the 2-surface. In this case, it's normal only to the 2-surface and this is sufficient. The general evolution can be split up into a normal and a tangential part, where the tangential part would contribute to a ``shift" of the 2-surface coordinates from one leaf to another. Even if the initially chosen hypersurface has no shift, a shift might develop in the course of the deformation \cite{booth2013}. Moreover, without spherical symmetry in our spacetime, there is no ``preferred" slicing and we shouldn't expect the marginally trapped tubes (MTTs)\footnote{MTTs are hypersurfaces foliated by marginally trapped surfaces.} to be unique. For a spherical symmetric spacetime, a spherical slicing yields a unique MTT \cite{booth2010} (See \cite{schnetter2006a} for a discussion on non-spherical slicing in Vaidya).
In our case, the deviation from spherical symmetry is due to the supertranslation field and we have chosen $f\ll m$. This amounts to only first-order corrective deviations from spherical symmetry. In fact, we see that the normal of the null surface and the radial null vector differ only in the $\theta$-component (shown here at order $\epsilon^1$):
\[
l^a=(1,2\dot m,0,0),\quad \Phi^{,a}=\left(1,2\dot m, \frac{\dot m f'}{2m^2},0\right).
\]
In the slowly evolving regime, we also have $\dot m\ll m$, implying that the $\theta$-component is extremely small.

\begin{figure*}
\begin{centering}
\includegraphics[width=\linewidth]{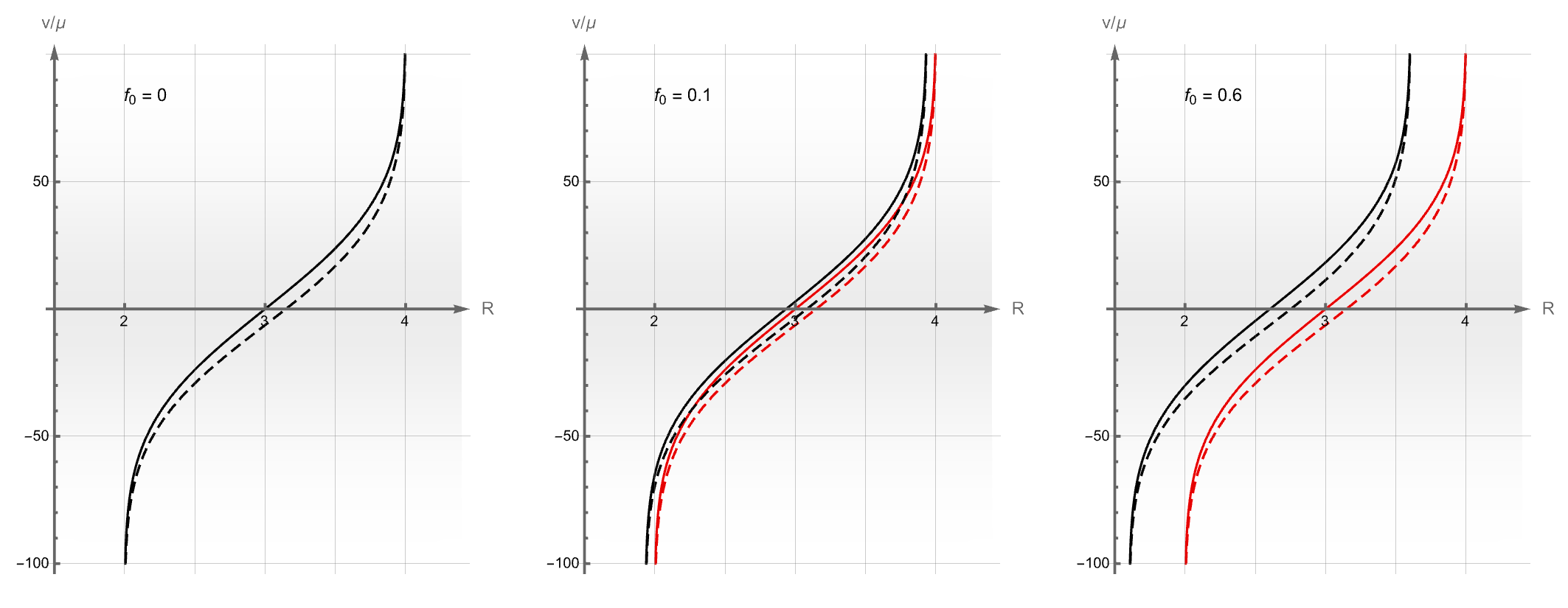}
\caption{The solid curves represent the FOTH, the dashed curves represent the null surface for $\theta=\pi/2$ (red) and $\theta=\pi/4$ (black). We see that for $f_0=0$, surfaces of different $\theta$ coincide, as expected. As we increase $f_0$, the separation between surfaces at different $\theta$ values increase. These results will of course vary depending on the ansatz for the supertranslation field chosen.}
\label{fig:stv-reh-2}
\end{centering}
\end{figure*}


The expansion of the radial null congruence at this surface is (at leading order in $\epsilon$)
\[
\Theta_{(l)}=\epsilon\frac{2\dot M}{\mu M}-\epsilon\frac{\dot M}{2\mu M^2}\Big(\mathscr{f}'\cot{\theta}+\mathscr{f''}\Big)
\]
which is again, positive - hence untrapped. This congruence, however, also possesses shear. The shear 2-tensor $\sigma^{(l)}_{AB}$ for this congruence is (at leading order in $\epsilon$):
\[
\sigma^{(l)}_{AB}=\epsilon\mu\dot M(\mathscr{f}'\cot{\theta}-\mathscr{f}'')\Big([d\theta]_A [d\theta]_B-\sin^2 \theta\;[d\phi]_A [d\phi]_B\Big)
\]
which clearly shows that the shear is entirely due to the supertranslation field.

The conditions for an SENS now includes the shear as well. We need to compute $\lie{l}\Theta_{(l)}$, $\kappa_{(l)}$ and $G_{ab}l^a l^b$, to leading order in $\epsilon$:
\begin{equation}
\begin{gathered}
\lie{l}\Theta_{(l)}=\frac{2\epsilon^2}{\mu^2 M^2}(M\ddot M-\dot M^2)\\+\frac{\epsilon^2}{2\mu^2 M^3}(\mathscr{f}'\cot{\theta}+\mathscr{f''})(2\dot M^2-M\ddot M)\\
\kappa_{(l)}=\frac{1}{4\mu M}\qquad G_{ab}l^a l^b=\epsilon\frac{\dot M}{2\mu^2 M^2}
\end{gathered}
\end{equation}
where we see that the supertranslation field does not appear at all in the leading order in $\epsilon$ in $\kappa_{(l)}$ and $G_{ab}l^a l^b$. They are exactly the same as in the Vaidya case.
$\|\sigma^{(l)}\|^2$ on the other hand, can be neglected, which we can confirm without explicit computation. This is because $\sigma^{(l)}_{AB}$ has $O(f)$ dependence and $\|\sigma^{(l)}\|^2$ will naturally have $O(f^2)$ dependence, terms which we will ignore in our formulation. Thus even in this case, the slow evolution conditions look similar to the spherically symmetric one.
Then, with $\epsilon\ll 1$ and the bounds on $f$ previously imposed, the slowly evolving conditions are satisfied.

\section{Vaidya BH in 5-D EGB gravity}\label{sec:vaidya-bh-in-5-d-egb-gravity}
In the 5-D Einstein-Gauss-Bonnet theory, a spherically symmetric asymptotically flat Vaidya-like solution can be generated  \cite{ghosh2008, boulware1985} for the field equation $G_{ab}+\alpha H_{ab}=8\pi T_{ab}$ with $H_{ab}$ signifies the contribution from the Gauss-Bonnet higher curvature part.  
\begin{equation}
\begin{split}
ds^2 &= -\triangle(v,r)dv^2+2dvdr+r^2 d \Omega_3^2\\
\triangle(v,r) &= 1+ \frac{r^2}{4\alpha}\left[1-\sqrt{1+\frac{8\alpha m(v)}{r^4}}\right]
\end{split}
\end{equation}
where $d\Omega_3^2=d \theta^2 + \sin^2{\theta}\,d \phi^2 + \sin^2{\theta}\sin^2{\phi}\,d \psi^2$, the line element of the $v,r$ constant 3-surface and $\alpha$ is the Gauss-Bonnet coupling constant. We consider $\alpha> 0$.

The stress-energy tensor for the Vaidya solution (note it does not include $\alpha$):
\begin{equation}
T_{ab}=\frac{3\dot m}{2r^3}\delta^v_a \delta^v_b
\end{equation}
Then to satisfy the null energy condition, we demand
\[
T_{ab}l^a l^b\geq 0 \implies \dot m \geq 0
\]
for an outgoing null vector $l^a$.
\subsection{Location of the FOTH}\label{sec:locating-the-foth}
We again focus on outgoing and ingoing radial null geodesics to determine the horizons. The tangent vectors to these are, much like the VRN case,
\begin{equation}
l^a\coloneqq\left(1,\frac{\triangle}{2},0,0,0\right),\quad n^a\coloneqq(0,-1,0,0,0).
\end{equation}
The corresponding expansions are
\[
\Theta_{(l)}=\frac{3\triangle}{2r},\quad \quad \Theta_{(n)}=-\frac{3}{r}
\]
Then the MTS are given, as usual, by $\Theta_{(l)}= 0 = \triangle$. This yields the solution
\begin{equation}
r_+(v)=\sqrt{m(v)-2\alpha}.
\end{equation}
Since $\Theta_{(n)}=-3/r < 0$, this is of the future type. We also calculate
\[
\lie{n}\Theta_{(l)}=\frac{3}{2r^2}\Big(\triangle-r\,\partial_r \triangle\Big)
\]
which, evaluated at the MTS, is
\[
\lie{n}\Theta_{(l)}=-\frac{3}{2\alpha+m}\cdot
\]
This is always negative, given our assumptions. Then this MTS is indeed the FOTH. It also naturally requires us to impose the restriction $m(v)>2\alpha$ to get physically meaningful solutions. For other types of collapsing BH in EGB theory one may see \cite{chatterjee2022} where the evolution of MTSs are studied.  The norm of the normal to the FOTH is given by
\[
g^{ab}\Phi_{,a}\Phi_{,b}=-\frac{\dot m}{r_+},
\]
which is timelike, given the energy condition holds.
\subsection{Location of the SENS}\label{sec:locating-the-sens}
Now we try to (perturbatively) find a null surface close to the FOTH. First, we transform to dimensionless quantities again, noting that $[\alpha]\sim [m]\sim [r^2]$.
\[
r\to \mu R,\;v\to L\mu V,\;m\to \mu^2 M,\;\alpha\to \mu^2 A.
\]
Since $\alpha$ is supposed to be fixed by theory, $A$ changes with the choice of $\mu$. Moreover, now only derivatives of $m$ acquire a factor of $\mu$:
\[
\dot m(v)=\odr{m(v)}{v}\to \frac{\mu^2}{\mu L}\odr{M(V)}{V}= \frac{\mu}{L}\dot M(V).
\]
The modification of the metric is as in the case of VRN or STV spacetimes.

The perturbative expansion is exactly as given in the VRN case, in terms of $\epsilon$. Again, we assume the hierarchy of derivatives: $R_+ \gtrsim \dot R_+ \gtrsim |\ddot R_+| \;\dots$ We plug this expansion into the nullity condition. Then the first order coefficient of $\epsilon$ is
\[
A_1\equiv \frac{(2A+M)\dot M}{2 R_+^3}\cdot
\]
Naturally, given all our assumptions, this is positive. In this case, the qualitative behaviour seen in the SENS plot is very similar to the one seen for the VRN metric in \cref{fig:vrn-seh}.
The other relevant results are:
\begin{equation}\label{sens-egb}
\begin{gathered}
\Theta_{(l)}=\frac{3\epsilon\dot M}{2\mu R_+^2},\quad \kappa_{(l)}=\frac{R_+}{\mu(2A+M)}\\
G_{ab}l^a l^b=\frac{3\epsilon\dot M}{2\mu^2 R_+(2A+M)}\\ 
\lie{l}\Theta_{(l)}=-\frac{3\epsilon^2}{2\mu^2 R_+^4}\Big(\dot M^2-R_+^2\ddot M\Big)
\end{gathered}
\end{equation}

The slowly evolving conditions in this case are
\[
\frac{1}{3}\Theta_{(l)}^2\ll G_{ab}l^a l^b,\quad \text{and}\quad\lie{l}\Theta_{(l)}\ll \kappa_{(l)}\Theta_{(l)}.
\]
We would also need to calculate the evolution parameter $C$ on the FOTH:
\begin{equation}
    C = \frac{\epsilon\dot M}{2R_+}
\end{equation}
and we require $C\ll 1$. This puts the restriction $\epsilon\dot M\ll 2R_+$, which implies how large the derivative is allowed to be depends on the size of the horizon. For instance, this assures that the first order coefficient $A_1$ with $A = 0$ does not diverge in the limit $M\to 0$. Then the slowly evolving conditions would hold, given $\epsilon\ll 1$.

\section{The laws of black hole mechanics for SENS}
One of the motivations behind the slowness conditions is that it allows to establish an approximate version of the first law of black hole mechanics to hold. When the slowness conditions are met one can write an approximate version of first law for SENS as \cite{booth2013}  
\[\kappa_{(l)}\Theta_{(l)}\approx \big(\|\sigma_{(l)}\|^2 + \mathcal{R}_{ab}l^a l^b\big),\]
where $q$ is the determinant of the metric $q_{AB}$. The interpretation of this expression has to be elaborated a little bit. It is apparent that the relation reminds us the Clausius statement of first law $T\delta S\approx\delta E$. The left hand side of the expression can easily be identified as a product of the surface gravity with rate of change of area of a slice of the SENS. The right hand side is the flux through the slice. The energy or flux terms are due to external matter as well as gravitational shear. The term $\mathcal{R}_{ab}l^al^b$ can be replaced by $G_{ab}l^al^b$ and eventually by $T_{ab}l^al^b$ via Einstein equation for general relativity. For the SENS, the surface gravity $\kappa_{(l)}$ coincides with that of FOTH to the leading order. Further it can be expressed as a slow variation from $\kappa_0$ that denotes the surface gravity for the equilibrium state (isolated horizon). Then the SENS satisfy an integrated first law of the form
\begin{equation}
    \kappa_{0}\dot{A}\approx \int_{S_v}\sqrt{q}\,dA\big(\|\sigma_{(l)}\|^2 + 8\pi T_{ab}l^a l^b\big),
\label{flaw1}\end{equation}
where the dot denotes derivative w.r.t.\ $v$, and $\sqrt{q}\,dA$ is the area element of a slice $S_v$. This form is quite similar to the Physical Process version of the first law \cite{wald1994}. Also it is known that spherically symmetric SENS would not have the shear term. The Vaidya like metricmetrics are examples of that. On the other hand the STV metric admits the shear term but the contribution of the squared shear can be ignored as it is produced to the second order in $\epsilon.$ The gravitational shear part however dominates for rotating solutions or whenever it is not induced by any perturbation.   For theories beyond GR like EGB gravity (with spherical metrics) we would get additional contribution in first law from the higher curvature terms. The energy flux would then have an additional $-\alpha H_{ab}l^al^b$ part. However, the apparent negative contribution does not spoil the positiveness of the flux as the $\mathcal{R}_{ab}l^al^b$ has already been shown to be positive for SENS in (\ref{sens-egb}). The null energy condition then ensures a positive flux across the slices $S_v$. 

We now can establish the area increase law for the SENS. This can be achieved by the use of Raychaudhuri equation for the null congruence generated by $l$:
\begin{equation}\label{re-scl}
    \kappa_{(l)}\Theta_{(l)}-\lie{l}\Theta_{(l)}=\frac{1}{d-1}\Theta_{(l)}^2+\|\sigma_{(l)}\|^2+\mathcal{R}_{ab}l^a l^b.
\end{equation}
When the slowness conditions hold, the expansion term of the right hand side can be neglected in comparison to the rest \cite{booth2013} Similarly the left hand side is dominated by the $\kappa_{(l)}$ term. Therefore for a positive surface gravity, the $\Theta_{(l)}$ must be positive when the geometry satisfies the NEC in GR. This must be true for any slice and therefore establishes the monotonicity of area of the SENS slices. 

This area increase law would be valid for the SENS we have considered here. For the EGB-Vaidya solution again the positiveness of the $G_{ab}l^al^b$ term ensures the positivity of the right hand side of (\ref{re-scl}). 


It might be worthwhile to note here that for higher curvature gravity theories area is not proportional to entropy. Therefore the second law of thermodynamics can't be extended as area increase law for such theories. For any general covariant theory, Wald's prescription provides a candidate for entropy \cite{wald1993}. For higher curvature theories a linearized second law has been proven in \cite{wall2015} adopting a picture where a stationary horizon is perturbed and becomes non-stationary. However, for such cases the entropy candidate should be chosen carefully as Wald entropy  might not be the correct candidate in dynamical situations \cite{iyer1994}. In such a scenario the second law of black hole mechanics can be established using Jacobson-Kang-Myers (JKM) entropy \cite{jacobson1994}. In the EGB theory the higher curvature contribution of the JKM entropy is the Ricci scalar of the $3$ dimensional spherical surface on the event horizon \cite{chatterjee2012,bhattacharjee2016}. Since SENS is a candidate for the event horizon and the slowly evolving regime can be regarded as a slight perturbation from stationarity, one can check if the JKM entropy increases on this surface or not. The change in entropy functional can thus be written as:
\begin{eqnarray}
    \odr{S}{\lambda}&=&\frac{1}{4}\frac{d}{d\lambda}\Big((1+2\alpha ^{(3)}\mathcal{R})\sqrt{q}\,dA\Big)\nonumber\\
    &=&\frac{1}{4}\sqrt{q}\,dA\Big(1+\frac{4\alpha}{r^2}\Big)\Theta_{(l)},\label{cinentropy}
\end{eqnarray}
where $\lambda$ parametrizes the horizon generators and we have inserted the value of Ricci scalar for a spherically symmetric $3$-surface $^{(3)}\mathcal{R}=6/r^2.$ The expression (\ref{cinentropy}) is clearly positive for $\alpha > 0$ as the expansion has already been proven positive in the area increase law. Hence the JKM entropy increases on the SENS as long as the slowness condition holds.

\section{Discussions}

In the pursuit of slowly evolving null surfaces in different spacetimes using radial null geodesics only, we found something unique in each to talk about. In all of these cases, the perturbative expansion allows us to analytically find a null surface that ``encapsulates" the FOTH and coincides well with the numerically determined event horizon in the slowly evolving regime. 

The charged Vaidya case is a simple demonstration of how the evolution scale of the spacetime is controlled not only by time derivatives of the mass and charge functions (which make the space dynamical in the first place) but also by the proximity of $q$ to $m$. Approaching this limit increasingly pushes our spacetime away from the slowly evolving regime, which is why we obtained a rough bound on the proximity. In a way, it is the relative scale of $q$ w.r.t $m$ that is crucial. Unless we stick to a scale, the perturbative coefficients become arbitrarily large and the expansion breaks down.

In the STV case, we broke free from spherical symmetry at a corrective level and discussed the consequences of evolution along a vector which is not normal to the FOTH. This introduced shear into the problem, which allows us to associate gravitational radiation with the SENS. We also obtained bounds on the magnitude of the supertranslation field in order for the black hole to be slowly evolving. Finally, for the Vaidya black hole in 5-D EGB gravity, we noted that there is no hard bound on $\dot m$, unlike in the simple 4-D Vaidya case - its magnitude is bounded by the size of the horizon.

We have shown the approximate versions of first and second laws of black hole mechanics can be recovered for SENS for the class of BH we have studied. The first law for SENS resemble with the physical process version of first laws for event horizon. Our approach can be extended for the spherically symmetric solutions of any dimension higher than $5$ for the EGB BH. It is intriguing to note, the JKM entropy increases on the SENS for EGB-Vaidya solution. This indicates the SENS indeed behaves like a weakly perturbed analogue of an event horizon. A solution independent proof of the area and entropy increase laws for higher curvature gravity theories would be worthy to attempt. It may be interesting to explore the membrane and fluid aspects of dynamical and slowly evolving horizons as the study may reveal useful facts of dynamical BH \cite{gourgoulhon2006}. This approach of slowly evolving horizons may further be compared with the recent results found on black hole thermodynamics beyond Einstein gravity \cite{bhattacharya2020a}.


\section{Acknowledgements}
The authors acknowledge useful discussions with Subhodeep Sarkar and Divyesh Solanki. S.B. thanks Ayan Chatterjee for helpful comments on various issues of the paper.
A.T. thanks Rounak Chatterjee for helpful insights. The research of S.B. is partially supported by SERB-DST through MATRICS grant MTR/2022/000170. 


\begin{thebibliography}{34}%
  \makeatletter
  \providecommand \@ifxundefined [1]{%
   \@ifx{#1\undefined}
  }%
  \providecommand \@ifnum [1]{%
   \ifnum #1\expandafter \@firstoftwo
   \else \expandafter \@secondoftwo
   \fi
  }%
  \providecommand \@ifx [1]{%
   \ifx #1\expandafter \@firstoftwo
   \else \expandafter \@secondoftwo
   \fi
  }%
  \providecommand \natexlab [1]{#1}%
  \providecommand \enquote  [1]{``#1''}%
  \providecommand \bibnamefont  [1]{#1}%
  \providecommand \bibfnamefont [1]{#1}%
  \providecommand \citenamefont [1]{#1}%
  \providecommand \href@noop [0]{\@secondoftwo}%
  \providecommand \href [0]{\begingroup \@sanitize@url \@href}%
  \providecommand \@href[1]{\@@startlink{#1}\@@href}%
  \providecommand \@@href[1]{\endgroup#1\@@endlink}%
  \providecommand \@sanitize@url [0]{\catcode `\\12\catcode `\$12\catcode
    `\&12\catcode `\#12\catcode `\^12\catcode `\_12\catcode `\%12\relax}%
  \providecommand \@@startlink[1]{}%
  \providecommand \@@endlink[0]{}%
  \providecommand \url  [0]{\begingroup\@sanitize@url \@url }%
  \providecommand \@url [1]{\endgroup\@href {#1}{\urlprefix }}%
  \providecommand \urlprefix  [0]{URL }%
  \providecommand \Eprint [0]{\href }%
  \providecommand \doibase [0]{https://doi.org/}%
  \providecommand \selectlanguage [0]{\@gobble}%
  \providecommand \bibinfo  [0]{\@secondoftwo}%
  \providecommand \bibfield  [0]{\@secondoftwo}%
  \providecommand \translation [1]{[#1]}%
  \providecommand \BibitemOpen [0]{}%
  \providecommand \bibitemStop [0]{}%
  \providecommand \bibitemNoStop [0]{.\EOS\space}%
  \providecommand \EOS [0]{\spacefactor3000\relax}%
  \providecommand \BibitemShut  [1]{\csname bibitem#1\endcsname}%
  \let\auto@bib@innerbib\@empty
  \bibitem [{\citenamefont {Bardeen}\ \emph {et~al.}(1973)\citenamefont
    {Bardeen}, \citenamefont {Carter},\ and\ \citenamefont
    {Hawking}}]{bardeen1973}%
    \BibitemOpen
    \bibfield  {author} {\bibinfo {author} {\bibfnamefont {J.~M.}\ \bibnamefont
    {Bardeen}}, \bibinfo {author} {\bibfnamefont {B.}~\bibnamefont {Carter}},\
    and\ \bibinfo {author} {\bibfnamefont {S.~W.}\ \bibnamefont {Hawking}},\
    }\bibfield  {title} {\bibinfo {title} {The four laws of black hole
    mechanics},\ }\href {https://doi.org/10.1007/BF01645742} {\bibfield
    {journal} {\bibinfo  {journal} {Commun.Math. Phys.}\ }\textbf {\bibinfo
    {volume} {31}},\ \bibinfo {pages} {161} (\bibinfo {year} {1973})}\BibitemShut
    {NoStop}%
  \bibitem [{\citenamefont {Hawking}(1975)}]{hawking1975}%
    \BibitemOpen
    \bibfield  {author} {\bibinfo {author} {\bibfnamefont {S.~W.}\ \bibnamefont
    {Hawking}},\ }\bibfield  {title} {\bibinfo {title} {Particle creation by
    black holes},\ }\href {https://doi.org/10.1007/BF02345020} {\bibfield
    {journal} {\bibinfo  {journal} {Commun.Math. Phys.}\ }\textbf {\bibinfo
    {volume} {43}},\ \bibinfo {pages} {199} (\bibinfo {year} {1975})}\BibitemShut
    {NoStop}%
  \bibitem [{\citenamefont {Wald}(2001)}]{wald2001}%
    \BibitemOpen
    \bibfield  {author} {\bibinfo {author} {\bibfnamefont {R.~M.}\ \bibnamefont
    {Wald}},\ }\bibfield  {title} {\bibinfo {title} {The {{Thermodynamics}} of
    {{Black Holes}}},\ }\href {https://doi.org/10.12942/lrr-2001-6} {\bibfield
    {journal} {\bibinfo  {journal} {Living Rev. in Rel.}\ }\textbf {\bibinfo
    {volume} {4}},\ \bibinfo {pages} {6} (\bibinfo {year} {2001})},\ \Eprint
    {https://arxiv.org/abs/gr-qc/9912119} {arXiv:gr-qc/9912119} \BibitemShut
    {NoStop}%
  \bibitem [{\citenamefont {Hayward}(1994)}]{hayward1994}%
    \BibitemOpen
    \bibfield  {author} {\bibinfo {author} {\bibfnamefont {S.~A.}\ \bibnamefont
    {Hayward}},\ }\bibfield  {title} {\bibinfo {title} {General {{Laws}} of
    {{Black-Hole Dynamics}}},\ }\href {https://doi.org/10.1103/PhysRevD.49.6467}
    {\bibfield  {journal} {\bibinfo  {journal} {Phys. Rev. D}\ }\textbf {\bibinfo
    {volume} {49}},\ \bibinfo {pages} {6467} (\bibinfo {year} {1994})},\ \Eprint
    {https://arxiv.org/abs/gr-qc/9303006} {arXiv:gr-qc/9303006} \BibitemShut
    {NoStop}%
  \bibitem [{\citenamefont {Ashtekar}\ \emph {et~al.}(2000)\citenamefont
    {Ashtekar}, \citenamefont {Beetle}, \citenamefont {Dreyer}, \citenamefont
    {Fairhurst}, \citenamefont {Krishnan}, \citenamefont {Lewandowski},\ and\
    \citenamefont {Wisniewski}}]{ashtekar2000}%
    \BibitemOpen
    \bibfield  {author} {\bibinfo {author} {\bibfnamefont {A.}~\bibnamefont
    {Ashtekar}}, \bibinfo {author} {\bibfnamefont {C.}~\bibnamefont {Beetle}},
    \bibinfo {author} {\bibfnamefont {O.}~\bibnamefont {Dreyer}}, \bibinfo
    {author} {\bibfnamefont {S.}~\bibnamefont {Fairhurst}}, \bibinfo {author}
    {\bibfnamefont {B.}~\bibnamefont {Krishnan}}, \bibinfo {author}
    {\bibfnamefont {J.}~\bibnamefont {Lewandowski}},\ and\ \bibinfo {author}
    {\bibfnamefont {J.}~\bibnamefont {Wisniewski}},\ }\bibfield  {title}
    {\bibinfo {title} {Generic {{Isolated Horizons}} and their
    {{Applications}}},\ }\href {https://doi.org/10.1103/PhysRevLett.85.3564}
    {\bibfield  {journal} {\bibinfo  {journal} {Phys. Rev. Lett.}\ }\textbf
    {\bibinfo {volume} {85}},\ \bibinfo {pages} {3564} (\bibinfo {year}
    {2000})},\ \Eprint {https://arxiv.org/abs/gr-qc/0006006}
    {arXiv:gr-qc/0006006} \BibitemShut {NoStop}%
  \bibitem [{\citenamefont {Ashtekar}\ and\ \citenamefont
    {Krishnan}(2003)}]{ashtekar2003}%
    \BibitemOpen
    \bibfield  {author} {\bibinfo {author} {\bibfnamefont {A.}~\bibnamefont
    {Ashtekar}}\ and\ \bibinfo {author} {\bibfnamefont {B.}~\bibnamefont
    {Krishnan}},\ }\bibfield  {title} {\bibinfo {title} {Dynamical {{Horizons}}
    and their {{Properties}}},\ }\href
    {https://doi.org/10.1103/PhysRevD.68.104030} {\bibfield  {journal} {\bibinfo
    {journal} {Phys. Rev. D}\ }\textbf {\bibinfo {volume} {68}},\ \bibinfo
    {pages} {104030} (\bibinfo {year} {2003})},\ \Eprint
    {https://arxiv.org/abs/gr-qc/0308033} {arXiv:gr-qc/0308033} \BibitemShut
    {NoStop}%
  \bibitem [{\citenamefont {Ashtekar}\ and\ \citenamefont
    {Krishnan}(2004)}]{ashtekar2004}%
    \BibitemOpen
    \bibfield  {author} {\bibinfo {author} {\bibfnamefont {A.}~\bibnamefont
    {Ashtekar}}\ and\ \bibinfo {author} {\bibfnamefont {B.}~\bibnamefont
    {Krishnan}},\ }\bibfield  {title} {\bibinfo {title} {Isolated and dynamical
    horizons and their applications},\ }\href
    {https://doi.org/10.12942/lrr-2004-10} {\bibfield  {journal} {\bibinfo
    {journal} {Living Rev. Relativ.}\ }\textbf {\bibinfo {volume} {7}},\ \bibinfo
    {pages} {10} (\bibinfo {year} {2004})},\ \Eprint
    {https://arxiv.org/abs/gr-qc/0407042} {arXiv:gr-qc/0407042} \BibitemShut
    {NoStop}%
  \bibitem [{\citenamefont {Thorne}\ \emph {et~al.}(1986)\citenamefont {Thorne},
    \citenamefont {Price},\ and\ \citenamefont {MacDonald}}]{thorne1986}%
    \BibitemOpen
    \bibfield  {author} {\bibinfo {author} {\bibfnamefont {K.~S.}\ \bibnamefont
    {Thorne}}, \bibinfo {author} {\bibfnamefont {R.~H.}\ \bibnamefont {Price}},\
    and\ \bibinfo {author} {\bibfnamefont {D.~A.}\ \bibnamefont {MacDonald}},\
    }\href@noop {} {\emph {\bibinfo {title} {Black Holes: {{The}} Membrane
    Paradigm}}}\ (\bibinfo {year} {1986})\BibitemShut {NoStop}%
  \bibitem [{\citenamefont {Booth}\ and\ \citenamefont
    {Fairhurst}(2004)}]{booth2004}%
    \BibitemOpen
    \bibfield  {author} {\bibinfo {author} {\bibfnamefont {I.}~\bibnamefont
    {Booth}}\ and\ \bibinfo {author} {\bibfnamefont {S.}~\bibnamefont
    {Fairhurst}},\ }\bibfield  {title} {\bibinfo {title} {The first law for
    slowly evolving horizons},\ }\href
    {https://doi.org/10.1103/PhysRevLett.92.011102} {\bibfield  {journal}
    {\bibinfo  {journal} {Phys. Rev. Lett.}\ }\textbf {\bibinfo {volume} {92}},\
    \bibinfo {pages} {011102} (\bibinfo {year} {2004})},\ \Eprint
    {https://arxiv.org/abs/gr-qc/0307087} {arXiv:gr-qc/0307087} \BibitemShut
    {NoStop}%
  \bibitem [{\citenamefont {Booth}\ and\ \citenamefont
    {Fairhurst}(2007)}]{booth2007}%
    \BibitemOpen
    \bibfield  {author} {\bibinfo {author} {\bibfnamefont {I.}~\bibnamefont
    {Booth}}\ and\ \bibinfo {author} {\bibfnamefont {S.}~\bibnamefont
    {Fairhurst}},\ }\bibfield  {title} {\bibinfo {title} {Isolated, slowly
    evolving, and dynamical trapping horizons: Geometry and mechanics from
    surface deformations},\ }\href {https://doi.org/10.1103/PhysRevD.75.084019}
    {\bibfield  {journal} {\bibinfo  {journal} {Phys. Rev. D}\ }\textbf {\bibinfo
    {volume} {75}},\ \bibinfo {pages} {084019} (\bibinfo {year} {2007})},\
    \Eprint {https://arxiv.org/abs/gr-qc/0610032} {arXiv:gr-qc/0610032}
    \BibitemShut {NoStop}%
  \bibitem [{\citenamefont {Booth}\ and\ \citenamefont
    {Martin}(2010)}]{booth2010}%
    \BibitemOpen
    \bibfield  {author} {\bibinfo {author} {\bibfnamefont {I.}~\bibnamefont
    {Booth}}\ and\ \bibinfo {author} {\bibfnamefont {J.}~\bibnamefont {Martin}},\
    }\bibfield  {title} {\bibinfo {title} {On the proximity of black hole
    horizons: Lessons from {{Vaidya}}},\ }\href
    {https://doi.org/10.1103/PhysRevD.82.124046} {\bibfield  {journal} {\bibinfo
    {journal} {Phys. Rev. D}\ }\textbf {\bibinfo {volume} {82}},\ \bibinfo
    {pages} {124046} (\bibinfo {year} {2010})},\ \Eprint
    {https://arxiv.org/abs/1007.1642} {arXiv:1007.1642 [gr-qc]} \BibitemShut
    {NoStop}%
  \bibitem [{\citenamefont {Booth}(2013)}]{booth2013}%
    \BibitemOpen
    \bibfield  {author} {\bibinfo {author} {\bibfnamefont {I.}~\bibnamefont
    {Booth}},\ }\bibfield  {title} {\bibinfo {title} {Spacetime near isolated and
    dynamical trapping horizons},\ }\href
    {https://doi.org/10.1103/PhysRevD.87.024008} {\bibfield  {journal} {\bibinfo
    {journal} {Phys. Rev. D}\ }\textbf {\bibinfo {volume} {87}},\ \bibinfo
    {pages} {024008} (\bibinfo {year} {2013})},\ \Eprint
    {https://arxiv.org/abs/1207.6955} {arXiv:1207.6955 [gr-qc, physics:hep-th]}
    \BibitemShut {NoStop}%
  \bibitem [{\citenamefont {Dreyer}\ \emph {et~al.}(2003)\citenamefont {Dreyer},
    \citenamefont {Krishnan}, \citenamefont {Schnetter},\ and\ \citenamefont
    {Shoemaker}}]{dreyer2003a}%
    \BibitemOpen
    \bibfield  {author} {\bibinfo {author} {\bibfnamefont {O.}~\bibnamefont
    {Dreyer}}, \bibinfo {author} {\bibfnamefont {B.}~\bibnamefont {Krishnan}},
    \bibinfo {author} {\bibfnamefont {E.}~\bibnamefont {Schnetter}},\ and\
    \bibinfo {author} {\bibfnamefont {D.}~\bibnamefont {Shoemaker}},\ }\bibfield
    {title} {\bibinfo {title} {Introduction to {{Isolated Horizons}} in
    {{Numerical Relativity}}},\ }\href
    {https://doi.org/10.1103/PhysRevD.67.024018} {\bibfield  {journal} {\bibinfo
    {journal} {Phys. Rev. D}\ }\textbf {\bibinfo {volume} {67}},\ \bibinfo
    {pages} {024018} (\bibinfo {year} {2003})},\ \Eprint
    {https://arxiv.org/abs/gr-qc/0206008} {arXiv:gr-qc/0206008} \BibitemShut
    {NoStop}%
  \bibitem [{\citenamefont {Bonnor}\ and\ \citenamefont
    {Vaidya}(1970)}]{bonnor1970}%
    \BibitemOpen
    \bibfield  {author} {\bibinfo {author} {\bibfnamefont {W.~B.}\ \bibnamefont
    {Bonnor}}\ and\ \bibinfo {author} {\bibfnamefont {P.~C.}\ \bibnamefont
    {Vaidya}},\ }\bibfield  {title} {\bibinfo {title} {Spherically symmetric
    radiation of charge in {{Einstein-Maxwell}} theory},\ }\href
    {https://doi.org/10.1007/BF00756891} {\bibfield  {journal} {\bibinfo
    {journal} {Gen Relat Gravit}\ }\textbf {\bibinfo {volume} {1}},\ \bibinfo
    {pages} {127} (\bibinfo {year} {1970})}\BibitemShut {NoStop}%
  \bibitem [{\citenamefont {Mishra}\ \emph {et~al.}(2019)\citenamefont {Mishra},
    \citenamefont {Chakraborty},\ and\ \citenamefont {Sarkar}}]{mishra2019}%
    \BibitemOpen
    \bibfield  {author} {\bibinfo {author} {\bibfnamefont {A.~K.}\ \bibnamefont
    {Mishra}}, \bibinfo {author} {\bibfnamefont {S.}~\bibnamefont
    {Chakraborty}},\ and\ \bibinfo {author} {\bibfnamefont {S.}~\bibnamefont
    {Sarkar}},\ }\bibfield  {title} {\bibinfo {title} {Understanding photon
    sphere and black hole shadow in dynamically evolving spacetimes},\ }\href
    {https://doi.org/10.1103/PhysRevD.99.104080} {\bibfield  {journal} {\bibinfo
    {journal} {Phys. Rev. D}\ }\textbf {\bibinfo {volume} {99}},\ \bibinfo
    {pages} {104080} (\bibinfo {year} {2019})},\ \Eprint
    {https://arxiv.org/abs/1903.06376} {arXiv:1903.06376 [gr-qc]} \BibitemShut
    {NoStop}%
  \bibitem [{\citenamefont {Nielsen}(2014)}]{nielsen2014}%
    \BibitemOpen
    \bibfield  {author} {\bibinfo {author} {\bibfnamefont {A.~B.}\ \bibnamefont
    {Nielsen}},\ }\bibfield  {title} {\bibinfo {title} {Revisiting {{Vaidya
    Horizons}}},\ }\href {https://doi.org/10.3390/galaxies2010062} {\bibfield
    {journal} {\bibinfo  {journal} {Galaxies}\ }\textbf {\bibinfo {volume} {2}},\
    \bibinfo {pages} {62} (\bibinfo {year} {2014})}\BibitemShut {NoStop}%
  \bibitem [{\citenamefont {Nielsen}\ and\ \citenamefont
    {Shoom}(2018)}]{nielsen2018}%
    \BibitemOpen
    \bibfield  {author} {\bibinfo {author} {\bibfnamefont {A.~B.}\ \bibnamefont
    {Nielsen}}\ and\ \bibinfo {author} {\bibfnamefont {A.~A.}\ \bibnamefont
    {Shoom}},\ }\bibfield  {title} {\bibinfo {title} {Conformal {{Killing}}
    horizons and their thermodynamics},\ }\href
    {https://doi.org/10.1088/1361-6382/aab505} {\bibfield  {journal} {\bibinfo
    {journal} {Class. Quantum Grav.}\ }\textbf {\bibinfo {volume} {35}},\
    \bibinfo {pages} {105008} (\bibinfo {year} {2018})}\BibitemShut {NoStop}%
  \bibitem [{\citenamefont {Kavanagh}\ and\ \citenamefont
    {Booth}(2006)}]{kavanagh2006}%
    \BibitemOpen
    \bibfield  {author} {\bibinfo {author} {\bibfnamefont {W.}~\bibnamefont
    {Kavanagh}}\ and\ \bibinfo {author} {\bibfnamefont {I.}~\bibnamefont
    {Booth}},\ }\bibfield  {title} {\bibinfo {title} {Spacetimes containing
    slowly evolving horizons},\ }\href
    {https://doi.org/10.1103/PhysRevD.74.044027} {\bibfield  {journal} {\bibinfo
    {journal} {Phys. Rev. D}\ }\textbf {\bibinfo {volume} {74}},\ \bibinfo
    {pages} {044027} (\bibinfo {year} {2006})},\ \Eprint
    {https://arxiv.org/abs/gr-qc/0603074} {arXiv:gr-qc/0603074} \BibitemShut
    {NoStop}%
  \bibitem [{\citenamefont {Strominger}(2018)}]{strominger2018}%
    \BibitemOpen
    \bibfield  {author} {\bibinfo {author} {\bibfnamefont {A.}~\bibnamefont
    {Strominger}},\ }\href {https://doi.org/10.48550/arXiv.1703.05448} {\bibinfo
    {title} {Lectures on the {{Infrared Structure}} of {{Gravity}} and {{Gauge
    Theory}}}} (\bibinfo {year} {2018}),\ \Eprint
    {https://arxiv.org/abs/1703.05448} {arXiv:1703.05448 [astro-ph,
    physics:gr-qc, physics:hep-ph, physics:hep-th, physics:math-ph]} \BibitemShut
    {NoStop}%
  \bibitem [{\citenamefont {Chu}\ and\ \citenamefont {Koyama}(2018)}]{chu2018}%
    \BibitemOpen
    \bibfield  {author} {\bibinfo {author} {\bibfnamefont {C.-S.}\ \bibnamefont
    {Chu}}\ and\ \bibinfo {author} {\bibfnamefont {Y.}~\bibnamefont {Koyama}},\
    }\bibfield  {title} {\bibinfo {title} {Soft {{Hair}} of {{Dynamical Black
    Hole}} and {{Hawking Radiation}}},\ }\href
    {https://doi.org/10.1007/JHEP04(2018)056} {\bibfield  {journal} {\bibinfo
    {journal} {J. High Energ. Phys.}\ }\textbf {\bibinfo {volume}
    {2018}}\bibfield  {number} {\bibinfo  {number} { (4)},\ \bibinfo {pages}
    {56}},\ }\Eprint {https://arxiv.org/abs/1801.03658} {arXiv:1801.03658 [gr-qc,
    physics:hep-th]} \BibitemShut {NoStop}%
  \bibitem [{\citenamefont {Sarkar}\ \emph {et~al.}(2022)\citenamefont {Sarkar},
    \citenamefont {Kumar},\ and\ \citenamefont {Bhattacharjee}}]{sarkar2022}%
    \BibitemOpen
    \bibfield  {author} {\bibinfo {author} {\bibfnamefont {S.}~\bibnamefont
    {Sarkar}}, \bibinfo {author} {\bibfnamefont {S.}~\bibnamefont {Kumar}},\ and\
    \bibinfo {author} {\bibfnamefont {S.}~\bibnamefont {Bhattacharjee}},\
    }\bibfield  {title} {\bibinfo {title} {Can we detect a supertranslated black
    hole?},\ }\href {https://doi.org/10.1103/PhysRevD.105.084001} {\bibfield
    {journal} {\bibinfo  {journal} {Phys. Rev. D}\ }\textbf {\bibinfo {volume}
    {105}},\ \bibinfo {pages} {084001} (\bibinfo {year} {2022})},\ \Eprint
    {https://arxiv.org/abs/2110.03547} {arXiv:2110.03547 [gr-qc, physics:hep-th]}
    \BibitemShut {NoStop}%
  \bibitem [{\citenamefont {Schnetter}\ and\ \citenamefont
    {Krishnan}(2006)}]{schnetter2006a}%
    \BibitemOpen
    \bibfield  {author} {\bibinfo {author} {\bibfnamefont {E.}~\bibnamefont
    {Schnetter}}\ and\ \bibinfo {author} {\bibfnamefont {B.}~\bibnamefont
    {Krishnan}},\ }\bibfield  {title} {\bibinfo {title} {Non-symmetric trapped
    surfaces in the {{Schwarzschild}} and {{Vaidya}} spacetimes},\ }\href
    {https://doi.org/10.1103/PhysRevD.73.021502} {\bibfield  {journal} {\bibinfo
    {journal} {Phys. Rev. D}\ }\textbf {\bibinfo {volume} {73}},\ \bibinfo
    {pages} {021502} (\bibinfo {year} {2006})},\ \Eprint
    {https://arxiv.org/abs/gr-qc/0511017} {arXiv:gr-qc/0511017} \BibitemShut
    {NoStop}%
  \bibitem [{\citenamefont {Ghosh}\ and\ \citenamefont
    {Deshkar}(2008)}]{ghosh2008}%
    \BibitemOpen
    \bibfield  {author} {\bibinfo {author} {\bibfnamefont {S.~G.}\ \bibnamefont
    {Ghosh}}\ and\ \bibinfo {author} {\bibfnamefont {D.~W.}\ \bibnamefont
    {Deshkar}},\ }\bibfield  {title} {\bibinfo {title} {Horizons of radiating
    black holes in {{Einstein}} gauss-bonnet gravity},\ }\href
    {https://doi.org/10.1103/PhysRevD.77.047504} {\bibfield  {journal} {\bibinfo
    {journal} {Phys. Rev. D}\ }\textbf {\bibinfo {volume} {77}},\ \bibinfo
    {pages} {047504} (\bibinfo {year} {2008})},\ \Eprint
    {https://arxiv.org/abs/0801.2710} {arXiv:0801.2710 [gr-qc]} \BibitemShut
    {NoStop}%
  \bibitem [{\citenamefont {Boulware}\ and\ \citenamefont
    {Deser}(1985)}]{boulware1985}%
    \BibitemOpen
    \bibfield  {author} {\bibinfo {author} {\bibfnamefont {D.~G.}\ \bibnamefont
    {Boulware}}\ and\ \bibinfo {author} {\bibfnamefont {S.}~\bibnamefont
    {Deser}},\ }\bibfield  {title} {\bibinfo {title} {String-{{Generated Gravity
    Models}}},\ }\href {https://doi.org/10.1103/PhysRevLett.55.2656} {\bibfield
    {journal} {\bibinfo  {journal} {Phys. Rev. Lett.}\ }\textbf {\bibinfo
    {volume} {55}},\ \bibinfo {pages} {2656} (\bibinfo {year}
    {1985})}\BibitemShut {NoStop}%
  \bibitem [{\citenamefont {Chatterjee}\ \emph {et~al.}(2022)\citenamefont
    {Chatterjee}, \citenamefont {Jaryal},\ and\ \citenamefont
    {Ghosh}}]{chatterjee2022}%
    \BibitemOpen
    \bibfield  {author} {\bibinfo {author} {\bibfnamefont {A.}~\bibnamefont
    {Chatterjee}}, \bibinfo {author} {\bibfnamefont {S.~C.}\ \bibnamefont
    {Jaryal}},\ and\ \bibinfo {author} {\bibfnamefont {A.}~\bibnamefont
    {Ghosh}},\ }\bibfield  {title} {\bibinfo {title} {Gravitational collapse in
    {{Einstein-Gauss-Bonnet}} gravity},\ }\href
    {https://doi.org/10.1103/PhysRevD.106.044049} {\bibfield  {journal} {\bibinfo
     {journal} {Phys. Rev. D}\ }\textbf {\bibinfo {volume} {106}},\ \bibinfo
    {pages} {044049} (\bibinfo {year} {2022})}\BibitemShut {NoStop}%
  \bibitem [{\citenamefont {Wald}(1994)}]{wald1994}%
    \BibitemOpen
    \bibfield  {author} {\bibinfo {author} {\bibfnamefont {R.~M.}\ \bibnamefont
    {Wald}},\ }\href@noop {} {\emph {\bibinfo {title} {Quantum {{Field Theory}}
    in {{Curved Spacetime}} and {{Black Hole Thermodynamics}}}}},\ Chicago
    {{Lectures}} in {{Physics}}\ (\bibinfo  {publisher} {{University of Chicago
    Press}},\ \bibinfo {address} {{Chicago, IL}},\ \bibinfo {year}
    {1994})\BibitemShut {NoStop}%
  \bibitem [{\citenamefont {Wald}(1993)}]{wald1993}%
    \BibitemOpen
    \bibfield  {author} {\bibinfo {author} {\bibfnamefont {R.~M.}\ \bibnamefont
    {Wald}},\ }\bibfield  {title} {\bibinfo {title} {Black {{Hole Entropy}} is
    {{Noether Charge}}},\ }\href {https://doi.org/10.1103/PhysRevD.48.R3427}
    {\bibfield  {journal} {\bibinfo  {journal} {Phys. Rev. D}\ }\textbf {\bibinfo
    {volume} {48}},\ \bibinfo {pages} {R3427} (\bibinfo {year} {1993})},\ \Eprint
    {https://arxiv.org/abs/gr-qc/9307038} {arXiv:gr-qc/9307038} \BibitemShut
    {NoStop}%
  \bibitem [{\citenamefont {Wall}(2015)}]{wall2015}%
    \BibitemOpen
    \bibfield  {author} {\bibinfo {author} {\bibfnamefont {A.~C.}\ \bibnamefont
    {Wall}},\ }\bibfield  {title} {\bibinfo {title} {A {{Second Law}} for
    {{Higher Curvature Gravity}}},\ }\href
    {https://doi.org/10.1142/S0218271815440149} {\bibfield  {journal} {\bibinfo
    {journal} {Int. J. Mod. Phys. D}\ }\textbf {\bibinfo {volume} {24}},\
    \bibinfo {pages} {1544014} (\bibinfo {year} {2015})},\ \Eprint
    {https://arxiv.org/abs/1504.08040} {arXiv:1504.08040 [gr-qc, physics:hep-th]}
    \BibitemShut {NoStop}%
  \bibitem [{\citenamefont {Iyer}\ and\ \citenamefont {Wald}(1994)}]{iyer1994}%
    \BibitemOpen
    \bibfield  {author} {\bibinfo {author} {\bibfnamefont {V.}~\bibnamefont
    {Iyer}}\ and\ \bibinfo {author} {\bibfnamefont {R.~M.}\ \bibnamefont
    {Wald}},\ }\bibfield  {title} {\bibinfo {title} {Some {{Properties}} of
    {{Noether Charge}} and a {{Proposal}} for {{Dynamical Black Hole Entropy}}},\
    }\href {https://doi.org/10.1103/PhysRevD.50.846} {\bibfield  {journal}
    {\bibinfo  {journal} {Phys. Rev. D}\ }\textbf {\bibinfo {volume} {50}},\
    \bibinfo {pages} {846} (\bibinfo {year} {1994})},\ \Eprint
    {https://arxiv.org/abs/gr-qc/9403028} {arXiv:gr-qc/9403028} \BibitemShut
    {NoStop}%
  \bibitem [{\citenamefont {Jacobson}\ \emph {et~al.}(1994)\citenamefont
    {Jacobson}, \citenamefont {Kang},\ and\ \citenamefont
    {Myers}}]{jacobson1994}%
    \BibitemOpen
    \bibfield  {author} {\bibinfo {author} {\bibfnamefont {T.}~\bibnamefont
    {Jacobson}}, \bibinfo {author} {\bibfnamefont {G.}~\bibnamefont {Kang}},\
    and\ \bibinfo {author} {\bibfnamefont {R.~C.}\ \bibnamefont {Myers}},\
    }\bibfield  {title} {\bibinfo {title} {On {{Black Hole Entropy}}},\ }\href
    {https://doi.org/10.1103/PhysRevD.49.6587} {\bibfield  {journal} {\bibinfo
    {journal} {Phys. Rev. D}\ }\textbf {\bibinfo {volume} {49}},\ \bibinfo
    {pages} {6587} (\bibinfo {year} {1994})},\ \Eprint
    {https://arxiv.org/abs/gr-qc/9312023} {arXiv:gr-qc/9312023} \BibitemShut
    {NoStop}%
  \bibitem [{\citenamefont {Chatterjee}\ and\ \citenamefont
    {Sarkar}(2012)}]{chatterjee2012}%
    \BibitemOpen
    \bibfield  {author} {\bibinfo {author} {\bibfnamefont {A.}~\bibnamefont
    {Chatterjee}}\ and\ \bibinfo {author} {\bibfnamefont {S.}~\bibnamefont
    {Sarkar}},\ }\bibfield  {title} {\bibinfo {title} {Physical process first law
    and increase of horizon entropy for black holes in {{Einstein-Gauss-Bonnet}}
    gravity},\ }\href {https://doi.org/10.1103/PhysRevLett.108.091301} {\bibfield
     {journal} {\bibinfo  {journal} {Phys. Rev. Lett.}\ }\textbf {\bibinfo
    {volume} {108}},\ \bibinfo {pages} {091301} (\bibinfo {year} {2012})},\
    \Eprint {https://arxiv.org/abs/1111.3021} {arXiv:1111.3021 [gr-qc,
    physics:hep-th]} \BibitemShut {NoStop}%
  \bibitem [{\citenamefont {Bhattacharjee}\ \emph {et~al.}(2016)\citenamefont
    {Bhattacharjee}, \citenamefont {Bhattacharyya}, \citenamefont {Sarkar},\ and\
    \citenamefont {Sinha}}]{bhattacharjee2016}%
    \BibitemOpen
    \bibfield  {author} {\bibinfo {author} {\bibfnamefont {S.}~\bibnamefont
    {Bhattacharjee}}, \bibinfo {author} {\bibfnamefont {A.}~\bibnamefont
    {Bhattacharyya}}, \bibinfo {author} {\bibfnamefont {S.}~\bibnamefont
    {Sarkar}},\ and\ \bibinfo {author} {\bibfnamefont {A.}~\bibnamefont
    {Sinha}},\ }\bibfield  {title} {\bibinfo {title} {Entropy functionals and
    c-theorems from the second law},\ }\href
    {https://doi.org/10.1103/PhysRevD.93.104045} {\bibfield  {journal} {\bibinfo
    {journal} {Phys. Rev. D}\ }\textbf {\bibinfo {volume} {93}},\ \bibinfo
    {pages} {104045} (\bibinfo {year} {2016})},\ \Eprint
    {https://arxiv.org/abs/1508.01658} {arXiv:1508.01658 [gr-qc, physics:hep-th]}
    \BibitemShut {NoStop}%
  \bibitem [{\citenamefont {Gourgoulhon}\ and\ \citenamefont
    {Jaramillo}(2006)}]{gourgoulhon2006}%
    \BibitemOpen
    \bibfield  {author} {\bibinfo {author} {\bibfnamefont {E.}~\bibnamefont
    {Gourgoulhon}}\ and\ \bibinfo {author} {\bibfnamefont {J.~L.}\ \bibnamefont
    {Jaramillo}},\ }\bibfield  {title} {\bibinfo {title} {Area evolution, bulk
    viscosity and entropy principles for dynamical horizons},\ }\href
    {https://doi.org/10.1103/PhysRevD.74.087502} {\bibfield  {journal} {\bibinfo
    {journal} {Phys. Rev. D}\ }\textbf {\bibinfo {volume} {74}},\ \bibinfo
    {pages} {087502} (\bibinfo {year} {2006})},\ \Eprint
    {https://arxiv.org/abs/gr-qc/0607050} {arXiv:gr-qc/0607050} \BibitemShut
    {NoStop}%
  \bibitem [{\citenamefont {Bhattacharya}\ \emph {et~al.}(2020)\citenamefont
    {Bhattacharya}, \citenamefont {Bhattacharyya}, \citenamefont {Dinda},\ and\
    \citenamefont {Kundu}}]{bhattacharya2020a}%
    \BibitemOpen
    \bibfield  {author} {\bibinfo {author} {\bibfnamefont {J.}~\bibnamefont
    {Bhattacharya}}, \bibinfo {author} {\bibfnamefont {S.}~\bibnamefont
    {Bhattacharyya}}, \bibinfo {author} {\bibfnamefont {A.}~\bibnamefont
    {Dinda}},\ and\ \bibinfo {author} {\bibfnamefont {N.}~\bibnamefont {Kundu}},\
    }\bibfield  {title} {\bibinfo {title} {An entropy current for dynamical black
    holes in four-derivative theories of gravity},\ }\href
    {https://doi.org/10.1007/JHEP06(2020)017} {\bibfield  {journal} {\bibinfo
    {journal} {J. High Energ. Phys.}\ }\textbf {\bibinfo {volume}
    {2020}}\bibfield  {number} {\bibinfo  {number} { (6)},\ \bibinfo {pages}
    {17}},\ }\Eprint {https://arxiv.org/abs/1912.11030} {arXiv:1912.11030 [gr-qc,
    physics:hep-th]} \BibitemShut {NoStop}%
  \end{thebibliography}

%

\end{document}